\documentclass[10pt,journal,epsfig]{IEEEtran}
\usepackage{graphicx}
\usepackage{epstopdf}
\usepackage{amssymb}
\usepackage{cite}
\usepackage{subfig}
\usepackage{amsmath}
\usepackage{algorithm}
\usepackage{algorithmic}
\usepackage{multirow}
\usepackage{xcolor}
\usepackage{stfloats}
\newtheorem{assumption}{Assumption}

\newtheorem{proposition}{Proposition}

\newcommand{\bm}{\boldsymbol}

\begin{document}

\title{Intelligent Reflecting Surface-Assisted NLOS Sensing With OFDM Signals}

\author{Jilin Wang, Jun Fang, Hongbin
Li,~\IEEEmembership{Fellow,~IEEE}, and Lei Huang
\thanks{Part of this work was submitted to EUSIPCO 2024. The current
work is a full version which includes theoretical analyses and
provides a more detailed algorithmic development.}
\thanks{Jilin Wang and Jun Fang are with the National Key Laboratory
of Wireless Communications, University of Electronic Science and
Technology of China, Chengdu 611731, China (e-mail: jilinwang@std.uestc.edu.cn, JunFang@uestc.edu.cn).}
\thanks{Hongbin Li is with the Department of Electrical and Computer Engineering,
Stevens Institute of Technology, Hoboken, NJ 07030, USA (e-mail:
Hongbin.Li@stevens.edu).}
\thanks{Lei Huang is with the State Key Laboratory of Radio
Frequency Heterogeneous Integration, Shenzhen University, Shenzhen 518060, China (e-mail: lhuang@szu.edu.cn).}
}

\maketitle
\begin{abstract}
This work addresses the problem of intelligent reflecting surface
(IRS) assisted target sensing in a non-line-of-sight (NLOS)
scenario, where an IRS is employed to facilitate the radar/access
point (AP) to sense the targets when the line-of-sight (LOS) path
between the AP and the target is blocked by obstacles. To sense
the targets, the AP transmits a train of uniformly-spaced
orthogonal frequency division multiplexing (OFDM) pulses, and then
perceives the targets based on the echoes from the
AP-IRS-targets-IRS-AP channel. To resolve an inherent scaling
ambiguity associated with IRS-assisted NLOS sensing, we propose a
two-phase sensing scheme by exploiting the diversity in the
illumination pattern of the IRS across two different phases.
Specifically, the received echo signals from the two phases are
formulated as third-order tensors. Then a canonical polyadic (CP)
decomposition-based method is developed to estimate each target's
parameters including the direction of arrival (DOA), Doppler shift
and time delay. Our analysis reveals that the proposed method
achieves reliable NLOS sensing using a modest quantity of
pulse/subcarrier resources. Simulation results are provided to
show the effectiveness of the proposed method under the
challenging scenario where the degrees-of-freedom provided by the
AP-IRS channel are not enough for resolving the scaling ambiguity.
\end{abstract}


\section{Introduction}
\subsection{Background}
Intelligent reflecting surface (IRS) has received a great amount
of attention in wireless communications due to its ability of
reconfiguring wireless propagation channels
\cite{8811733,9140329,8910627}. Specifically, IRS is made of a
newly developed metamaterial comprising a large number of
reconfigurable passive components. Through a smart controller, the
phase and amplitude of each unit on the IRS can be flexibly
adjusted. By properly designing the reflection coefficients, the
propagation environment can be customized to enhance/diminish
signals of interest. This allows for coherent or destructive
addition of reflected signals at the receiver, enabling passive
beamforming, increased spectral efficiency, interference
suppression, and other benefits
\cite{9326394,9226616,9234098,9927151}. In recent years, the
integration of wireless sensing as a new functionality into future
sixth-generation (6G) wireless networks has attracted increasing
research attention \cite{9540344,9737357,9893187,9893114}.
Wireless sensing typically involves extracting target information,
such as the angle and distance, through the line-of-sight (LOS) path
between the target and the wireless node. However, in some urban
scenarios, the targets of interest may be distributed in the
non-line-of-sight (NLOS) region of the wireless node, rendering
LOS path-based target sensing ineffective. To address this
challenge, IRS was introduced as an energy-efficient and
cost-effective anchor node with known locations, creating a
virtual LOS link between the sensing node and the target to
enhance performance
\cite{9508883,9454375,9732186,9361184,9647914,10149462,9724202,
9827797,10138058,9937163,10141975,10130707}.

\subsection{Related Works}
There have been some prior works investigating IRS-enabled
wireless sensing (i.e., NLOS detection/estimation)
\cite{9508883,9454375,9732186,9361184,9647914,10149462,9724202,9827797,10138058}
and IRS-assisted integrated sensing and communication (ISAC)
systems \cite{9937163,10141975,10130707}. For the IRS-aided NLOS
detection problem, the work \cite{9508883} developed a radar
equation for the IRS-aided NLOS scenario, and evaluated the
sensing performance in terms of signal-to-noise ratio (SNR) and
signal-to-clutter ratio (SCR). In \cite{9454375,9732186,9361184},
an IRS-aided multi-input multi-output (MIMO) radar detection
problem was considered, in which the IRS is placed in the vicinity
of radar transmitter (or receiver) to help illuminate (observe)
prospective targets. A generalized likelihood ratio test (GLRT)
detector was derived and the IRS phase shifts were optimized to
maximize the probability of detection given a fixed false alarm
probability. An IRS-assisted radar system for target surveillance
in a cluttered environment was studied in \cite{9647914}, where
the active beamformer at the radar transmitter and the passive
phase-shift matrices at IRSs are jointly optimized to maximize the
minimum target illumination power at multiple target locations.
The moving target detection problem in a multi-IRS-aided OFDM
radar system was considered in \cite{10149462}, where the authors
derived a bi-quadratic program which jointly designs the OFDM
signal and IRS phase shifts to optimize the target detection
performance.

In addition to detection, the estimation problem was studied for
IRS-aided NLOS sensing systems. The work \cite{9724202} considered
an IRS-self-sensing architecture, where an IRS controller is
employed to transmit probing signals, and dedicated sensors are
installed at the IRS for location/angle estimation based on the
echo signals via the BS-IRS-target-IRS sensor link and the
BS-target-IRS sensor link. An IRS-enabled pulse-Doppler radar
system was considered in \cite{9827797}, where the minimum
variance for the best linear unbiased estimator (BLUE) of the
target back-scattering coefficient is derived, and then the IRS
phase shifts were optimized by minimizing the mean squared error
of estimated target parameter. Moreover, the work \cite{10138058}
examined the estimation of the DOA in an IRS-enabled NLOS sensing
system, where the transmit beamformer at the AP and the passive
beamformer at the IRS were jointly optimized by minimizing the
Cram\'{e}r-Rao bound (CRB). It is noted in \cite{10138058} that an
inherent scaling ambiguity exists in IRS-assisted NLOS sensing
when the rank of the AP-IRS channel matrix is equal to one. This
is because at least two degrees-of-freedom (DoFs) are required to
identify both the complex path gain and the angular parameter of
the target, otherwise the scaling ambiguity arises
\cite{10422881}. To resolve the inherent scaling ambiguity, the
work \cite{10138058} needs that the AP-IRS channel matrix contains
at least two prominent singular values. Such a requirement,
however, may not be satisfied in practice. Specifically, for
IRS-assisted sensing, in order to compensate for the path loss
caused by multiple reflections, the IRS is usually located within
the sight of the AP and the AP-IRS channel is dominated by the LOS
path, in which case the work \cite{10138058} will experience a
substantial amount of performance degradation. For this reason, it
holds practical significance to study the scenario where the
AP-IRS channel is dominated by the LOS path.

\begin{figure}
    \centering
    {\includegraphics[width=.93\linewidth]{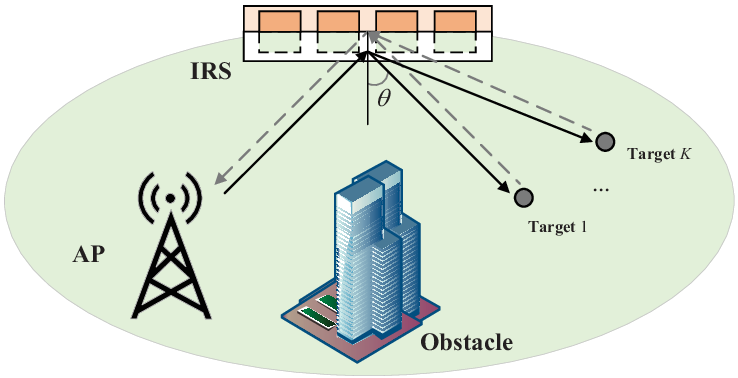}}
    \caption{System model of IRS-assisted sensing.}
    \label{system_model}
 \end{figure}

\subsection{Our Contributions}
In this paper, we consider the problem of target parameter
estimation via an IRS-assisted sensing system. The AP transmits a
train of uniformly-spaced OFDM pulses, and then perceives the
targets based on the echo signal from the AP-IRS-targets-IRS-AP
channel. To resolve the scaling ambiguity inherent in IRS-assisted
sensing, we, in this paper, propose a two-phase sensing method,
where the entire sensing cycle consists of two phases, and each
phase is assigned an individual IRS-phase-shift profile. By
utilizing the diversity of the IRS illumination pattern across two
phases, the received OFDM signals in two phases are represented by
two third-order tensors, and a CP decomposition-based method is
developed to uniquely identify the DOAs, time delays, and Doppler
shifts of the targets even when there is only a single dominant
path between the AP and the IRS. Additionally, a theoretical
analysis is presented to provide a performance bound for the
proposed sensing system. Simulation results demonstrate that the
proposed method achieves an estimation accuracy that is close to
the CRB, thereby validating the effectiveness of the proposed
method.

In addition to the ability of resolving the inherent scaling
ambiguity, our work presents some other advantages over
\cite{10138058}. First of all, the work \cite{10138058} only
studied the problem of DOA estimation, whereas our proposed method
can identify not only the DOA, but also the distance and the
Doppler shift parameters of the targets. Secondly, the work
\cite{10138058} considered only a single target scenario, and it
is difficult to extend the proposed maximum likelihood estimator
(MLE) to multi-target scenarios. As a comparison, our proposed
method can handle multiple targets simultaneously.

The remainder of this paper is organized as follows. Section
\ref{Formulation} introduces the system model as well as the
signal model of the proposed IRS-assisted NLOS sensing system.
Section \ref{ProposedMethod} develops a two-phase sensing scheme,
based on which the CP formulation, uniqueness conditions and CP
decomposition are discussed. Section \ref{ParaEst} discuss how to
estimate the target parameters from the estimated factor matrices.
Section \ref{CRBAnalysis} presents the CRB analysis for the
considered estimation problem. Simulation results are presented in
Section \ref{Simulations}, followed by concluding remarks in
Section \ref{Conclusion}.

\textit{Notations:} In this paper, scalars, column vectors,
matrices and tensors are denoted by italic, lowercase boldface,
uppercase boldface and  calligraphic boldface letters,
respectively. The symbols $(\cdot)^*$, $(\cdot)^T$, $(\cdot)^H$,
$(\cdot)^{-1}$, $(\cdot)^{\dagger}$ denote the conjugate,
transpose, conjugate transpose, inverse and pseudo-inverse,
respectively. $\Vert\cdot\Vert_2$ and $\Vert\cdot\Vert_F$ denote
the 2-norm and Frobenius norm, respectively.
$\mathrm{diag}(\bm{a})$ denotes a diagonal matrix whose main
diagonal elements are the elements of $\bm{a}$. $\bm{I}_M$ denotes
the identity matrix of size $M$. $[\bm{a}]_{i}$, $[\bm{A}]_{i,l}$,
$[\bm{A}]_{i,:}$, $[\bm{A}]_{:,l}$ denote the $i$th element of
$\bm{a}$, the $(i,l)$th element of $\bm{A}$, the $i$th row of
$\bm{A}$ and the $l$th column of $\bm{A}$, respectively.
$\mathrm{rank}(\bm{A})$ and $k_{\bm{A}}$ denote the rank and
Kruskal-rank of $\bm{A}$, respectively. $\otimes$, $\odot$,
$\circledast$ and $\circ$ denote the Kronecker, Khatri-Rao,
Hadamard and outer products, respectively. $j$ denotes the
imaginary unit. $\Re\{\cdot\}$ and $\Im\{\cdot\}$ denote the real
and imaginary parts of a complex number, respectively.

\section{Problem Formulation}
\label{Formulation}
\subsection{System Model}
Consider an IRS-assisted wireless sensing (i.e., radar) system,
where the LOS path between the radar/access point
and the target is blocked by obstacles (see
Fig. \ref{system_model}). The access point (AP) transmits a sensing
signal and then perceives the targets based on the echo signal
propagating through the AP-IRS-targets-IRS-AP channel. Suppose the
AP is equipped with a uniform linear array (ULA) of $M$ antennas,
and the IRS is equipped with a ULA of $N$ reflecting elements. We
assume that there are $K$ targets located in the area that are
illuminated by the IRS.
\begin{figure}[htbp]
    \centering
    {\includegraphics[width=.35\linewidth]{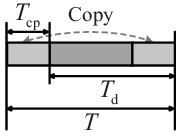}}
    \caption{One complete OFDM block.}
    \label{OFDM_BLOCK}
 \end{figure}

Let $\bm{x}(t)\in\mathbb{C}^{M}$ denote the transmitted signal,
and $\bm{G}\in\mathbb{C}^{N\times M}$ denote the channel matrix
from the AP to the IRS. Since the locations of the AP and the IRS
are pre-determined, we assume that the channel matrix $\bm{G}$ is
known \emph{a priori}. Each reflecting element of the IRS can
independently reflect the incident signal with a reconfigurable
phase shift. Define $\vartheta _n\in[0,2\pi]$ as the phase shift
associated with the $n$th reflecting element of the IRS. Also,
define the phase shift matrix of the IRS as
\begin{align}
    \bm{\Phi}=\mathrm{diag}(e^{j\vartheta _1},\cdots,e^{j\vartheta _n})\in\mathbb{C}^{N\times N}
\end{align}
Let $\theta$ denote a target's DOA with respect to the IRS. The
corresponding steering vector at the IRS can be written as
\begin{align}
\bm{a}(\theta)=\frac{1}{\sqrt{N}}[1\phantom{0}e^{j2\pi
\frac{d\mathrm{sin}(\theta)}{\lambda}}\phantom{0}
\cdots\phantom{0} e^{j2\pi
\frac{(N-1)d\mathrm{sin}(\theta)}{\lambda}}]^{T}
\label{IRS_steering}
\end{align}
where $d$ denotes the spacing between any two adjacent reflection
elements, and $\lambda$ is the wavelength of the carrier signal.
For the $k$th target, the cascaded IRS-target-IRS channel can be
written as
\begin{equation}
    \setlength{\abovedisplayskip}{3pt}
    \setlength{\belowdisplayskip}{3pt}
\begin{aligned}
\bm{H}_k=\tilde{\alpha}_k \bm{a}(\theta_k)\bm{a}^{T}(\theta_k)
\end{aligned}
\end{equation}
where $\tilde{\alpha}_k\in\mathbb{C}$ is used to characterize the
round-trip path loss as well as the radar cross section (RCS)
coefficient of the $k$th target. Define
$\bm{H}\triangleq\sum_{k=1}^{K}\bm{H}_k$. In this paper, we
consider the challenging scenario where the AP-IRS channel is
rank-one or approximately rank-one, i.e., $\text{rank}(\bm{G})=1$.
Nevertheless, as discussed later in this paper, our proposed
algorithm can be readily adapted to the less challenging scenario
where the rank of the AP-IRS channel is greater than one.

\subsection{Signal Model}
\subsubsection{Transmit signal model}
In a coherent processing interval (CPI), the AP transmits a train
of $P$ uniformly-spaced OFDM pulses. In each pulse, the AP
transmits one OFDM block and then receives the echo form potential
targets. Suppose there are $L$ orthogonal subcarriers in each
block and the subcarrier spacing is set as $\Delta
f=1/T_{\text{d}}$. The duration of one block is
$T=T_{\text{cp}}+T_{\text{d}}$, where $T_{\text{cp}}$ is the
length of the cyclic prefix and $T_{\text{d}}$ is the duration of
an OFDM symbol. The cyclic prefix is a replica of the end part of
the OFDM symbol (see Fig. \ref{OFDM_BLOCK}). Note in
communications, the duration of the cyclic prefix,
$T_{\text{cp}}$, should be larger than the time dispersion in a
radio channel with multipath propagation in order to avoid the
inter-symbol interference (ISI). While in this paper, the length
of cyclic prefix determines the maximum sensing distance from the
IRS to the target, which will be elaborated later.
\begin{figure}[htbp]
    \centering
    {\includegraphics[width=1\linewidth]{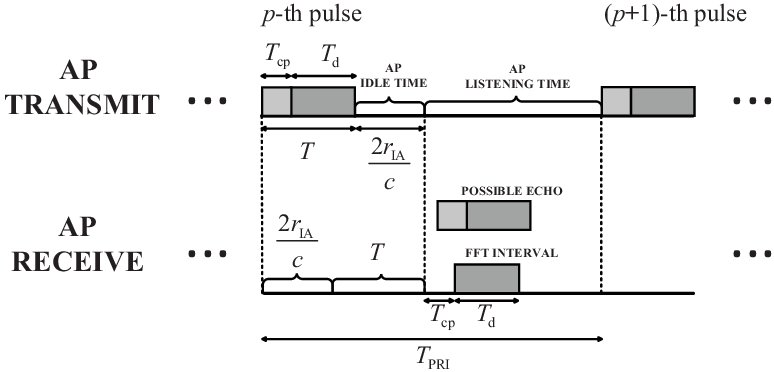}}
    \caption{A schematic of signal transmission in one pulse repetition interval.}
    \label{Signal_transmission}
 \end{figure}
Define $T_{\text{PRI}}$ as the pulse repetition interval (PRI).
The baseband signal in the $p$th pulse can be expressed as
\begin{align}
s_p(t) = \sum\limits_{l=1}^{L}\beta_{l}e^{j2\pi l\Delta ft}\xi
(t-pT_{\text{PRI}})
\label{baseband_tx}
\end{align}
where $pT_{\text{PRI}}\leq t\leq pT_{\text{PRI}}+T$, $\xi(t)$ is
the rectangular function that takes $1$ for $t\in[0,T]$ and $0$
otherwise \cite{5393298}, and $\beta_l$ is the unit-energy
modulated symbol which satisfies $|\beta_l|^2=1,\forall l$. For
such a signal, it can be readily verified that the cyclic prefix
part is a repetition of the end part of the OFDM block for any
$T_{\text{cp}}=\mu T, 1>\mu>0$. Also, for simplicity, we assume
$\beta_l=\beta,\forall l$ in this paper. Suppose we use an
individual transmit beamforming vector $\bm{w}_p\in\mathbb{C}^{M}$
to transmit the $p$th pulse. Then the transmitted signal can be
expressed as
\begin{align}
    \bm{x}_p(t) = \sqrt{P_t}\bm{w}_ps_p(t)\mathrm{exp}(j2\pi f_ct)
\end{align}
where $P_t$ denotes the transmit power and $f_c$ denotes the carrier
frequency.

\subsubsection{Received signal model}
Assume that the $k$th target is located at a distance of $R_k$
meters (m) from the IRS and the target is moving towards the IRS
with a radial velocity of $v_k$ (m/s). After transmitting the $p$th
pulse, the AP starts to listen to its echo signal after a duration of
$2r_{\text{IA}}/c$ seconds, where $r_{\text{IA}}$ denotes the
distance between the AP and the IRS. Such a duration is used as a
guard interval to avoid the interference signal directly reflected
from the IRS (see Fig. \ref{Signal_transmission}). This guard
interval also determines the minimum sensing distance from the
target to the IRS, which will be elaborated in the next section.
Also, we make the following assumption in order to acquire the
complete echo signal reflected from the targets.

\begin{assumption}
The echo signals from all potential targets are assumed to lie
within the interval of
$[2r_{\text{IA}}/c+T,2r_{\text{IA}}/c+2T+T_{\text{cp}}]$.
\label{assump1}
\end{assumption}

To process the received signal, a Fourier transform operation is
performed over the interval
$[2r_{\text{IA}}/c+T+T_{\text{cp}},2r_{\text{IA}}/c+2T]$. When
Assumption 1 is satisfied, it means that the earliest possible
echo signal reflected by a potential target will be received over
the interval $[2r_{\text{IA}}/c+T,2r_{\text{IA}}/c+2T]$, and the
latest possible echo signal reflected by a potential target will
be received within the interval
$[2r_{\text{IA}}/c+T+T_{\text{cp}},2r_{\text{IA}}/c+2T+T_{\text{cp}}]$.
Since the cyclic prefix part is a replica of the end part of the
OFDM block, the interval
$[2r_{\text{IA}}/c+T+T_{\text{cp}},2r_{\text{IA}}/c+2T]$ contains
each target's complete echo signal, in which case no information
will be lost. In other words, performing the Fourier transform
over the interval
$[2r_{\text{IA}}/c+T+T_{\text{cp}},2r_{\text{IA}}/c+2T]$ suffices
to retrieve the complete information of any echo signal, provided
that \textit{Assumption \ref{assump1}} is met.

Based on \textit{Assumption \ref{assump1}}, the pulse repetition
interval needs to satisfy $T_{\text{PRI}}\geq
2r_{\text{IA}}/c+2T+T_{\text{cp}}$. Since the AP operates in a
listening mode within the interval
$[2r_{\text{IA}}/c+T,T_{\text{PRI}}]$, the received echo signal
only contains signals reflected by targets. Thus, for the $p$th
pulse, the received signal at the $m$th antenna of the AP can be
written as
\begin{align}
    \tilde{y}_{p,m}(t) = \sum_{k=1}^{K}\bm{g}_m^T\bm{\Phi
}^T\bm{H}_k\bm{\Phi }\bm{G}\bm{x}_p(t-\tau_{p,k}) +
\tilde{n}_{p,m}(t)
\end{align}
where $\bm{g}_m$ is the $m$th column of $\bm{G}$,
$\tau_{p,k}=\frac{2(R_k+r_{\text{IA}}-v_k p T_{\text{PRI}})}{c}$
is the round-trip time delay associated with the $k$th target, $c$
is the speed of light and $\tilde{n}_{p,m}(t)$ is the additive
Gaussian noise. For notational simplicity, we define
$\tau_k\triangleq\frac{2R_k}{c}$,
$\nu_k\triangleq\frac{2v_kf_c}{c}$ and
$\tau_0\triangleq\frac{2r_{\text{IA}}}{c}$. We have
$\tau_{p,k}=\tau_k+\tau_0-{\nu_k p T_{\text{PRI}}}/{f_c}$.

After removing the carrier frequency, the baseband signal can be
written as (\ref{baseband_rx}) shown at the top of this page,
\begin{figure*}[htp]
    \begin{align}
    \bar{y}_{p,m}(t)
&=\sum_{k=1}^{K}\sqrt{P_t}\tilde{\alpha}_k\bm{g}_m^T\bm{\Phi}^T
\bm{a}(\theta_k)\bm{a}^{T}(\theta_k)\bm{\Phi}\bm{G}\bm{w}_p
s_p(t-\tau_{p,k})e^{-j2\pi f_c\tau_{p,k}}+\bar{n}_{p,m}(t) \nonumber \\
    &=\sum_{k=1}^{K}\sqrt{P_t}\tilde{\alpha}_k\bm{g}_m^T\bm{\Phi}^T\bm{a}(\theta_k)
    \bm{a}^{T}(\theta_k)\bm{\Phi}\bm{G}\bm{w}_p s_p(t-\tau_{p,k})
    e^{j2\pi pT_{\text{PRI}}\nu_k}e^{-j2\pi f_c(\tau_{k}+\tau_0)} + \bar{n}_{p,m}(t) \nonumber\\
    &=\sum_{k=1}^{K}{\bar{\alpha}}_k\bm{g}_m^T\bm{\Phi }^T
    \bm{a}(\theta_k)\bm{a}^{T}(\theta_k)\bm{\Phi}\bm{G}\bm{w}_p s_p(t-\tau_{p,k})
    e^{j2\pi pT_{\text{PRI}}\nu_k}+ \bar{n}_{p,m}(t) \nonumber\\
    &=\sum_{k=1}^{K}{\bar{\alpha}}_kb_m(\theta_k)z_{p}(\theta_k,\nu_k)s_p(t-\tau_{p,k})+\bar{n}_{p,m}(t)
    \label{baseband_rx}
\end{align}
\hrulefill
\end{figure*}
where
$\bar{\alpha}_k\triangleq\sqrt{P_t}\tilde{\alpha}_ke^{-j2\pi
f_c(\tau_k+\tau_0)}$,
 $b_m(\theta_k)\triangleq \bm{g}_m^T\bm{\Phi
}^T\bm{a}(\theta_k)$,
 $z_{p}(\theta_k,\nu_k)\triangleq
\bm{a}^{T}(\theta_k)\bm{\Phi }\bm{G}\bm{w}_pe^{j2\pi
pT_{\text{PRI}}\nu_k}$
and $\bar{n}_{p,m}(t)$ is the baseband noise.

Taking the Fourier transform of the received $p$th pulse baseband
signal over the interval
$[2r_{\text{IA}}/c+T+T_{\text{cp}},2r_{\text{IA}}/c+2T]$ (note
that $\tau_0=2r_{\text{IA}}/c$), the signal associated with the
$l$th subcarrier is given by
\begin{align}
    \tilde{y}_{p,m}[l] &= \int_{pT_{\text{PRI}}+\tau_0+T+T_{\text{cp}}}^{pT_{\text{PRI}}+
    \tau_0+2T}\bar{y}_{p,m}(t)e^{-j2\pi l\Delta ft} \,dt
    \label{fft_baseband}
\end{align}

\begin{figure*}[hbp]
    \hrulefill
    \begin{align}
\tilde{y}_{p,m}[l] &=\int_{pT_{\text{PRI}}+\tau_0+T+T_{\text{cp}}}^{pT_{\text{PRI}}+
\tau_0+2T}e^{-j2\pi
l\Delta ft} \sum_{k=1}^{K}{\bar{\alpha}}_k
b_m(\theta_k)z_{p}(\theta_k,\nu_k)\sum \limits_{q=1}^{L}\beta
e^{j2\pi q\Delta f(t-\tau_{p,k})} \,dt +
n_{p,m}[l] \nonumber\\
&=\sum_{k=1}^{K}{\bar{\alpha}}_kb_m(\theta_k)z_{p}(\theta_k,\nu_k)
\int_{pT_{\text{PRI}}+\tau_0+T+T_{\text{cp}}}^{pT_{\text{PRI}}+
\tau_0+2T}e^{-j2\pi
l\Delta ft} \sum\limits_{q=1}^{L}\beta e^{j2\pi q\Delta
f(t-\tau_{k}-\tau_{0})} e^{j2\pi
\frac{q\Delta f}{f_c}pT_{\text{PRI}}\nu_k}\,dt+ n_{p,m}[l] \nonumber\\
&\overset{(a)}\approx\sum_{k=1}^{K}{\bar{\alpha}}_kb_m(\theta_k)z_{p}(\theta_k,\nu_k)
\int_{pT_{\text{PRI}}+\tau_0+T+T_{\text{cp}}}^{pT_{\text{PRI}}+
\tau_0+2T}e^{-j2\pi
l\Delta ft} \sum\limits_{q=1}^{L}\beta e^{j2\pi q\Delta
f(t-\tau_{k}-\tau_{0})} \,dt+
n_{p,m}[l] \nonumber\\
&\overset{(b)}=\beta
T_{\text{d}}\sum_{k=1}^{K}{\bar{\alpha}}_kb_m(\theta_k)
z_{p}(\theta_k,\nu_k)e^{-j2\pi l\Delta f(\tau_k+\tau_{0})}+ n_{p,m}[l]
    \label{fft_result}
    \end{align}
\end{figure*}
Plugging (\ref{baseband_tx}) and (\ref{baseband_rx}) into
(\ref{fft_baseband}), we have (\ref{fft_result}) shown at the
bottom of this page, where the approximation $(a)$ follows from
the fact that the bandwidth of the baseband signal is far less
than the carrier frequency, i.e., $L\Delta f\ll f_c$, and $(b)$ is
due to the subcarrier orthogonality \cite{9420261}, i.e.
\begin{align}
    \int_{pT_{\text{PRI}}+\tau_0+T+T_{\text{cp}}}^{pT_{\text{PRI}}+
    \tau_0+2T} e^{j2\pi
(l\Delta f - q\Delta f)t} \,dt=T_{\text{d}}\delta(q\Delta f
-l\Delta f) \label{sub_orth}
\end{align}
Define $\alpha_k\triangleq \bar{\alpha}_k\beta T_{\text{d}}$, and
ignore the common phase term $\tau_0$ in (\ref{fft_result}) as
this term is known \emph{a priori}, we have
\begin{align}
y_{p,m}[l]=&\sum_{k=1}^{K}{\alpha}_k
b_m(\theta_k) z_{p}(\theta_k,\nu_k)e^{-j2\pi l\Delta f\tau_k}+
n_{p,m}[l]
\label{eqn1}
\end{align}
where
\begin{align}
   n_{p,m}[l] =\int_{pT_{\text{PRI}}+\tau_0+T+T_{\text{cp}}}^{pT_{\text{PRI}}+
   \tau_0+2T}\bar{n}_{p,m}(t)e^{-j2\pi l\Delta ft} \,dt
\end{align}
It is assumed that $n_{p,m}[l]$ is a complex Gaussian variable
with zeros mean and variance $\sigma^2$, i.e.,
$n_{p,m}[l]\sim\mathcal{CN}(0,\sigma^2)$.

\begin{figure*}[htp]
    \centering
    {\includegraphics[width=1\linewidth]{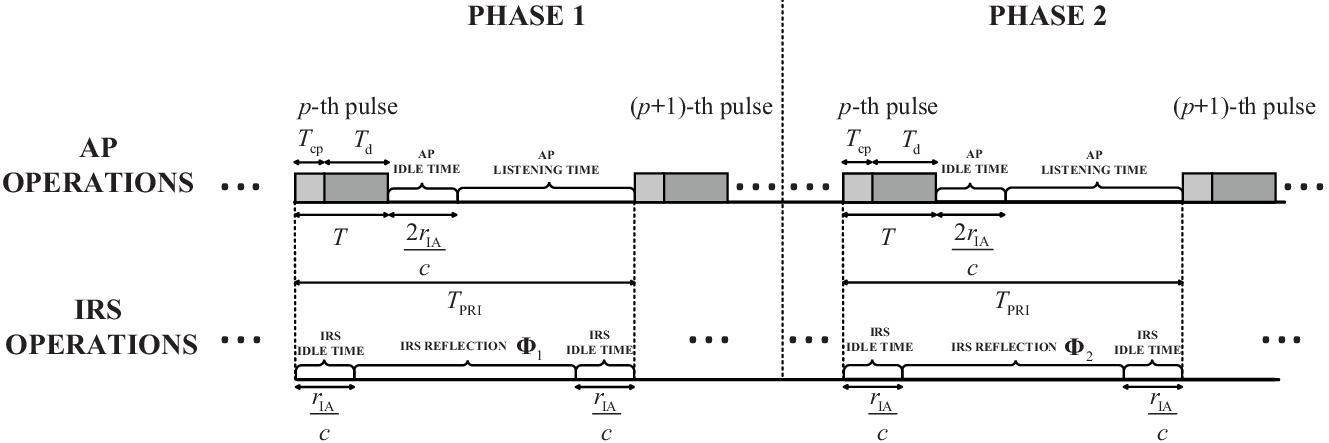}}
    \caption{A schematic of signal transmission for the two-phase NLOS sensing scheme.} \label{Scheme}
  \end{figure*}

\section{Proposed Sensing Scheme and CP Decomposition}
\label{ProposedMethod}

\subsection{Two-Phase Sensing Scheme}
The proposed two-phase sensing scheme is illustrated in Fig.
\ref{Scheme}, in which the entire sensing cycle is divided into
two phases, say, phase $1$ and phase $2$, and each of them is
assigned an individual IRS phase-shift profile. In each phase, the
AP transmits $P$ pulses in total. The pulse repetition interval is
$T_{\text{PRI}}$, and the interval between the $P$th pulse in
phase $1$ and the first pulse in phase $2$ is also set to
$T_{\text{PRI}}$.

In the following, we first analyze the minimum sensing distance
$R_{\text{min}}$, the maximum sensing distance $R_{\text{max}}$
and the maximum unambiguous velocity $v_{\text{max}}$ with respect
to the IRS.

\subsubsection{Minimum Sensing Distance}
To avoid the collision with the self-interference and the signal reflected directly from the
IRS, the reception is inhibited for the AP during the interval
$[pT_{\text{PRI}}, pT_{\text{PRI}}+2r_{\text{IA}}/c+T]$.
Accordingly, we can determine the minimum sensing distance from
the target to the IRS as
\begin{equation}
\begin{aligned}
R_{\text{min}}=\frac{c T}{2}
\end{aligned}
\end{equation}

\subsubsection{Maximum Sensing Distance}
According to Assumption 1, we know that the latest possible echo
signal reflected by a potential target will be received within the
interval
$[2r_{\text{IA}}/c+T+T_{\text{cp}},2r_{\text{IA}}/c+2T+T_{\text{cp}}]$.
As a result, the maximum sensing distance with respect to the IRS
is given by
\begin{equation}
\begin{aligned}
    R_{\text{max}}\leq \frac{c(T+T_{\text{cp}})}{2}
\end{aligned}
\end{equation}

\subsubsection{Unambiguous Velocity}
The maximum unambiguous velocity characterizes the maximum
detectable radial velocity of a target with respect to the IRS.
The radial velocity can be uniquely determined if there is no
phase ambiguity in $e^{j2\pi T_{\text{PRI}}\nu_k}$, i.e.,
$T_{\text{PRI}}\nu_k\leq 1$. Recalling
$\nu_k\triangleq\frac{2v_kf_c}{c}$, the maximum unambiguous
velocity can be given as
\begin{equation}
\begin{aligned}
    v_{\text{max}}\leq \frac{c}{2f_cT_{\text{PRI}}}
\end{aligned}
\end{equation}
\subsection{Tensor Representation}
Based on the above two-phase sensing scheme, we now show how to
formulate the received signals into tensors. Specifically, let
$\bm{\Phi}_1$ and $\bm{\Phi}_2$, respectively, denote the IRS
phase shift matrices employed in phase $1$ and phase $2$. Define
\begin{align}
  b_{i,m}(\theta_k)&\triangleq \bm{g}_m^T\bm{\Phi
}_i^T\bm{a}(\theta_k) \\
 z_{i,p}(\theta_k,\nu_k)&\triangleq
\bm{a}^{T}(\theta_k)\bm{\Phi }_i\bm{G}\bm{w}_pe^{j2\pi
pT_{\text{PRI}}\nu_k}
\end{align}
where $i\in\{1,2\}$. We first consider the signals received in
phase $1$. For each subcarrier $l$, stacking the received echo
signal from all $P$ pulses and all $M$ antennas, we can construct
a matrix $\bm{Y}_1(l)\in\mathbb{C}^{P\times M}$, with its
$(p,m)$th entry denoted by $[\bm{Y}_1(l)]_{p,m}=y^1_{p,m}[l]$ and
given as (\ref{eqn1})
\begin{equation}
\begin{aligned}
y^1_{p,m}[l] =& \sum_{k=1}^{K}{\alpha}_k b_{1,m}(\theta_k)\nonumber \\
&\times z_{1,p}(\theta_k,\nu_k)e^{-j2\pi l\Delta f\tau_k}+
n_{1,p,m}[l]
\end{aligned}
\end{equation}
Consequently, we have
\begin{equation}
\begin{aligned}
\bm{Y}_1(l)= \sum_{k=1}^{K}\alpha_kf_l(\tau_k)
\bm{z}_{1}(\theta_k,\nu_k)\bm{b}_1^H(\theta_k)+ \bm{N}_{1,l}
\end{aligned}
\end{equation}
where $\bm{z}_1(\theta_k,\nu_k)
\triangleq[z_{1,1}(\theta_k,\nu_k)\phantom{0}\cdots\phantom{0}
z_{1,P}(\theta_k,\nu_k)]^{T}\in\mathbb{C}^{P}$,
$\bm{b}_1(\theta_k)\triangleq[b_{1,1}(\theta_k)\phantom{0}\cdots\phantom{0}
b_{1,M}(\theta_k)]\in\mathbb{C}^{M}$, and $f_l(\tau_k)\triangleq
e^{-j2\pi l\Delta f\tau_k}$.

Now consider the signals received in phase $2$. In phase 1, the
distance between the $k$th target and the IRS is denoted as $R_k$.
When it comes to phase 2, the distance between the target and the
IRS has changed to $\tilde{R}_k=R_k+\Delta r_k$, where $\Delta
r_k=v_k P T_{\text{PRI}}$ is the shift of distance during the time
interval between two phases. For a typical sensing scenario,
suppose the velocity of the target is $v=120$km/h, the number of
pulses is set to $P=100$ and the PRI is set to
$T_{\text{PRI}}=10\mu\text{s}$. We have $\Delta r_k= vP
T_{\text{PRI}}\approx 0.03$m. This distance shift generally has a
very slight influence on the distance-dependent path loss and the
target's DOA with respect to IRS. Therefore, in phase $2$, it is
reasonable to assume that the path loss $\alpha$, the target's DOA
$\theta$, as well as the time delay $\tau$ remain the same as in
phase $1$. The received signals in phase $2$ can be written as
\begin{align}
y^2_{p,m}[l]
=&\sum_{k=1}^{K}{\alpha}_k b_{2,m}(\theta_k)\nonumber \\
&\times z_{2,p}(\theta_k,\nu_k)e^{-j2\pi l\Delta f\tau_k}+ n_{2,p,m}[l]
\end{align}
During phase $2$, let $\bm{Y}_2(l)\in\mathbb{C}^{P\times M}$
denote the matrix constructed by stacking the received echo signal
from all $P$ pulses and all $M$ antennas for each subcarrier $l$.
We have
\begin{align}
\bm{Y}_2(l)= \sum_{k=1}^{K}\alpha_kf_l({\tau}_k)
\bm{z}_{2}(\theta_k,\nu_k)\bm{b}_2^H(\theta_k)+ \bm{N}_{2,l}
\end{align}
where
$\bm{z}_2(\theta_k,\nu_k)\triangleq[z_{2,1}(\theta_k,\nu_k)\phantom{0}\cdots\phantom{0}
z_{2,P}(\theta_k,\nu_k)]^{T}\in\mathbb{C}^{P}$,
$\bm{b}_2(\theta_k)
\triangleq[b_{2,1}(\theta_k)\phantom{0}\cdots\phantom{0}
b_{2,M}(\theta_k)]\in\mathbb{C}^{M}$, $f_l(\tau_k)=e^{-j2\pi
l\Delta f\tau_k}$ and $\bm{N}_{2,l}$ is the additive Gaussian
noise.

For phase $i$, by concatenating the received signals across $L$
subcarriers, we can naturally obtain a third-order tensor
$\bm{\mathcal{Y}}_i\in\mathbb{C}^{P\times M\times L}$, with its
$(p,m,l)$th entry given by $[\bm{Y}_i(l)]_{p,m}$, whose three
modes respectively stand for the AP's antennas, the pulses and the
subcarriers. Note that each slice of the tensor
$\bm{\mathcal{Y}}_i$ is $\bm{Y}_i(l)$, which is a weighted sum of
a common set of rank-one outer products. Therefore the tensor
$\bm{\mathcal{Y}}_i$ admits a CP decomposition as
\begin{align}
\bm{\mathcal{Y}}_i =
\sum_{k=1}^{K}\bm{z}_i(\theta_k,\nu_k)\circ\bm{b}_i(\theta_k)\circ\alpha_k\bm{f}(\tau_k)
+ \bm{\mathcal{N}}_i \label{CPform}
\end{align}
where $\circ$ denotes the outer product, and we have
\begin{align}
\bm{f}(\tau_k)&\triangleq [e^{-j2\pi\Delta
f\tau_k}\phantom{0}\cdots\phantom{0}
e^{-j2\pi L\Delta f\tau_k}]^{T} \\
\bm{b}_i(\theta)&=\bm{G}^T\bm{\Phi}_{i}\bm{a}(\theta)\\
\bm{z}_i(\theta,\nu)&=(\bm{W}^T\bm{G}^T\bm{\Phi}_{i}\bm{a}(\theta))\circledast(\bm{{d}}(\nu))
\end{align}
in which
$\bm{W}\triangleq[\bm{w}_1\phantom{0}\cdots\phantom{0}\bm{w}_P]\in\mathbb{C}^{M\times
P}$, $\bm{{d}}(\nu)\triangleq[e^{j2\pi
T_{\text{PRI}}\nu}\phantom{0}\cdots\phantom{0} e^{j2\pi
PT_{\text{PRI}}\nu}]^T\in\mathbb{C}^{P}$ and $\circledast$ denotes the
Hadamard product.

Define
\begin{align}
    \bm{A}_i&\triangleq[\bm{z}_i(\theta_1,\nu_1)\phantom{0}
    \cdots\phantom{0}\bm{z}_i(\theta_K,\nu_K)]\in\mathbb{C}^{P\times K}\\
    \bm{B}_i&\triangleq[\bm{b}_i(\theta_1)\phantom{0}\cdots\phantom{0}
    \bm{b}_i(\theta_K)]\in\mathbb{C}^{M\times K}\\
    \bm{C}&\triangleq[\alpha_1\bm{f}(\tau_1)\phantom{0}\cdots\phantom{0}
    \alpha_K\bm{f}(\tau_K)]\in\mathbb{C}^{L\times K}
\end{align}
Here $\{\bm{A}_i,\bm{B}_i,\bm{C}\}$ are the factor matrices of the
tensor $\bm{\mathcal{Y}}_i$, where $i\in\{1,2\}$. We see that the
factor matrices contain information about the DOAs, Doppler shifts
and time delays of the targets. Leveraging the two phases'
observations, we, in the following, develop a two-stage method
which first estimates the factor matrices of the tensor
$\bm{\mathcal{Y}}_i$ and then jointly recovers the target's DOAs,
time delays and Doppler shifts based on the estimated factor
matrices. Before proceeding, we first discuss the uniqueness of CP
decomposition as it plays a crucial role in the identifiability of
the proposed method.

\subsection{Identifiability Condition}
A well-known condition to ensure the uniqueness of CP
decomposition is Kruskal's condition
\cite{harshman1972determination,kruskal1977three,stegeman2007kruskal},
i.e.
\begin{align}
k_{\boldsymbol{A}^{\left(1\right)}} +
k_{\boldsymbol{A}^{\left(2\right)}} +
k_{\boldsymbol{A}^{\left(3\right)}} \ge 2R + 2
\end{align}
where ${\boldsymbol{A}^{\left(1\right)}} \in \mathbb{C}^{I \times
R}$, ${\boldsymbol{A}^{\left(2\right)}} \in\mathbb{C}^ {{J \times
R}}$ and ${\boldsymbol{A}^{\left(3\right)}} \in\mathbb{C} {^{K
\times R}}$ are factor matrices associated with the third-order
tensor ${\boldsymbol{\cal X}}\in\mathbb{C}{^{I \times J \times
K}}$. $k_{\boldsymbol{A}}$ denotes the k-rank of a matrix
$\boldsymbol{A}$, which is defined as the largest value of
$k_{\boldsymbol{A}}$ such that every subset of
$k_{\boldsymbol{A}}$ columns of the matrix $\boldsymbol{A}$ is
linearly independent.

For our problem, recall that
$\boldsymbol{b}_{i}(\theta)=\boldsymbol{G}^T\boldsymbol{\Phi}_i\boldsymbol{a}(\theta)$,
and $\boldsymbol{G}$ is a rank-one matrix. Writing
$\boldsymbol{G}=\sigma\boldsymbol{u}\boldsymbol{v}^T$, we have
$\boldsymbol{b}_{i}(\theta)=c_i(\theta)\boldsymbol{v}$, where
$c_i(\theta)=\sigma\boldsymbol{u}^T\boldsymbol{\Phi}_i\boldsymbol{a}(\theta)$
is a scalar. Therefore, we can see that columns of the factor
matrix $\bm{B}_i$ are linearly dependent. Thus, we have
$k_{\bm{B}_i}=1$. In this case, even $\bm{A}_i$ and $\bm{C}$ are
full k-rank, the Kruskal's condition cannot be satisfied.

Meanwhile, we notice that the factor matrix $\boldsymbol{C}$ has a
Vandermonde structure. Previous studies found that, when one of
the factor matrices, say $\boldsymbol{A}^{(3)}$ has a Vandermonde
structure, the uniqueness of the CP decomposition can be
guaranteed if the following conditions are satisfied
\cite{6573422,9049103}:
\begin{align}
  \left\{\begin{array}{l}
    \operatorname{rank}\big(\underline{\boldsymbol{A}}^{(3)} \odot \boldsymbol{A}^{(2)}\big)=R \\
    \operatorname{rank}\big(\boldsymbol{A}^{(1)}\big)=R
    \end{array}\right.
\label{scp_condition}
\end{align}
where $\underline{\boldsymbol{A}}$ represents a sub-matrix of
$\boldsymbol{A}$ that is obtained by removing the bottom row of
$\boldsymbol{A}$, and $\odot$ denotes the Khatri-Rao product.

From the above condition (\ref{scp_condition}), we know that for
each $i$, if
\begin{align}
\left\{\begin{array}{l}
    \operatorname{rank}\left(\underline{\boldsymbol{C}} \odot \boldsymbol{B}_i\right)=K \\
    \operatorname{rank}\left(\boldsymbol{A}_i\right)=K
    \end{array}\right.
\label{eqn2}
\end{align}
then the CP decomposition of $\bm{\mathcal{Y}}_i$ is essentially
unique.

Note that different targets usually have different distances from
the IRS, $\bm{C}$ is thus a Vandermonde matrix with distinct
generators $\{e^{-j2\pi\Delta f\tau_k}\}$. According to
\cite{6573422}, we can arrive at
$\text{rank}(\underline{\bm{C}}\odot {\bm{B}_i})=K$ even if matrix
$\bm{B}_i$ has redundant columns, provided that $(L-1)M\geq K$.

Recall
$\boldsymbol{b}_{i}(\theta)=\boldsymbol{G}^T\boldsymbol{\Phi}_i\boldsymbol{a}(\theta)
=c_i(\theta)\boldsymbol{v}$. Therefore the column of the factor
matrix $\boldsymbol{A}_i$ can be expressed as
\begin{align}
\bm{z}_{i}(\theta,\nu)&=(\bm{W}^{T}\boldsymbol{G}^T
\boldsymbol{\Phi}_i\boldsymbol{a}(\theta))\circledast (\bm{d}(\nu)) \nonumber\\
&=(\bm{W}^T\boldsymbol{b}_{i}(\theta))\circledast(\bm{d}(\nu))\nonumber\\
&=(c_i(\theta)\bm{W}^T\boldsymbol{v})\circledast(\bm{d}(\nu))
\label{eqn3}
\end{align}
In practice, different targets are usually associated with
different Doppler shifts. From (\ref{eqn3}), we know that
$\bm{A}_i$ is obtained from a Vandermonde matrix multiplied
columnwisely by a same vector with different scaling factors.
Hence for a generic $\bm{W}^T\boldsymbol{v}$, $\bm{A}_i$ is full
rank and we have $\mathrm{rank}(\bm{A}_i)=\min\{P,K\},\forall i$.

In summary, we have the following proposition concerning the
uniqueness of the CP decomposition.
\begin{proposition}
Assume that the delay and Doppler shift parameters associated with
different targets are different. The uniqueness condition of CP
decomposition can be guaranteed almost surely when both
$(L-1)M\geq K$ and $P\geq K$ are satisfied. \label{prop1}
\end{proposition}

\subsection{CP Decomposition}
We now discuss how to perform CP decomposition by utilizing the
Vandermonde structure of the factor matrix. Such a method was
originally proposed in \cite{6573422}. To make the paper
self-contained, we provide a brief description of the CP
decomposition method.

To ease our presentation, we drop the subscript $i$ in the tensor
and the associated factor matrices. The mode-$1$ unfolding of
$\boldsymbol{\mathcal{Y}}$ can be written as
\begin{align}
  \bm{\mathcal{Y}}_{(1)}^{T}=(\bm{C}\odot \bm{B})\bm{A}^T +  \bm{\mathcal{N}}_{(1)}^{T}
  \label{mode1}
\end{align}
Ignoring the noise, we can compute the truncated singular value
decomposition (SVD) of the noiseless
$\bm{\mathcal{Y}}_{(1)}^{T}\in\mathbb{C} ^{LM\times P}$ as
\begin{equation}
\begin{aligned}
  \bm{\mathcal{Y}}_{(1)}^{T}=\bm{U}\bm{\Sigma}\bm{V}^H
\end{aligned}
\end{equation}
where $\bm{U} \in \mathbb{C}^{LM\times K}$, $\bm{\Sigma}\in
\mathbb{C}^{K\times K}$ and $\bm{V}\in \mathbb{C}^{P\times K}$. If
the uniqueness condition (\ref{eqn2}) is satisfied, there exists a
nonsingular matrix $\boldsymbol{R}\in\mathbb{C}^{K\times K}$ such
that
\begin{equation}
\begin{aligned}
  \boldsymbol{U}\boldsymbol{R}=\boldsymbol{C}\odot \boldsymbol{B}  \label{Gen_B}
\end{aligned}
\end{equation}
Define
$\bm{U}_{1}=[\bm{U}]_{1:(L-1)M,:}\in\mathbb{C}^{(L-1)M\times K}$
and $\bm{U}_{2}=[\bm{U}]_{M+1:LM,:}\in\mathbb{C}^{(L-1)M\times
K}$. We have
\begin{equation}
\begin{aligned}
  \boldsymbol{U}_1\boldsymbol{R}=\underline{\boldsymbol{C}}\odot \boldsymbol{B}\label{s1}  \\
  \boldsymbol{U}_2\boldsymbol{R}=\overline{\boldsymbol{C}}\odot \boldsymbol{B}
\end{aligned}
\end{equation}
where $\overline{\boldsymbol{C}}$ represents a submatrix of
$\boldsymbol{C}$ by removing the top row of $\boldsymbol{C}$, and
$\underline{\boldsymbol{C}}$ represents a submatrix of
$\boldsymbol{C}$ by removing the bottom row of $\boldsymbol{C}$.
Utilizing the Vandermonde structure of $\boldsymbol{C}$, we have
\begin{equation}
\begin{aligned}
  \left(\underline{\boldsymbol{C}}\odot \boldsymbol{B} \right)\boldsymbol{T}=
  \overline{\boldsymbol{C}}\odot \boldsymbol{B}
   \label{s2}
\end{aligned}
\end{equation}
in which $\boldsymbol{T}=\mathrm{diag}(t_{1},\cdots,t_{K})$ and
$t_{k}\triangleq e^{-j2\pi\vartriangle f \tau_{k}}$. Combining
(\ref{s1})--(\ref{s2}), we obtain
\begin{equation}
\begin{aligned}
  \boldsymbol{U}_{2}\boldsymbol{R}=\boldsymbol{U}_{1}\boldsymbol{R}\boldsymbol{T}\label{s3}
\end{aligned}
\end{equation}
According to (\ref{eqn2}) and (\ref{Gen_B}), both
$\boldsymbol{U}_{1}$ and $\boldsymbol{U}_{2}$ are full column
rank. Hence we can rewrite (\ref{s3}) as
\begin{align}
  \boldsymbol{U}_{1}^{\dagger}\boldsymbol{U}_{2}=\boldsymbol{R}\boldsymbol{T}\boldsymbol{R}^{-1}
\end{align}
Thus, we can perform the eigenvalue decomposition (EVD) of
$\boldsymbol{U}_{1}^{\dagger}\boldsymbol{U}_{2}$ to estimate
$\boldsymbol{T}$ and the associated generators
$\{t_{k}\}^{K}_{k=1}$. We can reconstruct the columns $\{\bm{
\hat{c}}_{k}\}$ of $\bm{\hat{C}}$ by
\begin{align}
\bm{\hat{c}}_{k}=[ \hat{t}_{k}\quad \hat{t}_{k}^2 \quad \cdots
\quad \hat{t}_{k}^L ]
\label{EstGen}
\end{align}
Based on (\ref{Gen_B}) and the reconstructed $\bm{\hat{C}}$, the column of the factor matrix
$\bm{\hat{B}}$ can be estimated as
\begin{align}
  \bm{ \hat{b}}_{k}=\left(\frac{\bm{ \hat{c}}_{k}^{H}}
  {\bm{ \hat{c}}_{k}^{H}\bm{ \hat{c}}_{k}}
  \otimes \boldsymbol{I}_{M}\right)\boldsymbol{U}[\bm{R}]_{:,k}
\end{align}
Finally, given $\hat{\boldsymbol{B}}$ and $\bm{\hat{C}}$, the
factor matrix ${\boldsymbol{A}}$ can be estimated as
\begin{equation}
\begin{aligned}
  \bm{\hat{A}}=\bm{\mathcal{Y}}_{(1)}\left(
  \left(\bm{\hat{C}}\odot\bm{\hat{B}}\right)^{T}\right)^{\dagger}
\end{aligned}
\end{equation}
After we obtain the estimated factor matrices $\bm{\hat{A}}$,
$\bm{\hat{B}}$ and $\bm{\hat{C}}$, we, in Section \ref{ParaEst},
discuss how to extract the sensing parameters from the estimated
factor matrices.

\section{Target Parameter Estimation}
\label{ParaEst}
After CP decomposition, we now have access to the
estimated factor matrices $\{
\bm{\hat{A}}_i,\bm{\hat{B}}_i,\bm{\hat{C}}_i\}$, in which
$i\in\{1,2\}$. Note that for both phases, the factor matrix
$\bm{C}_i$ remains the same, i.e. $\bm{C}_1=\bm{C}_2=\bm{C}$. Due
to the inherent permutation and scaling ambiguities, the estimated
factor matrices are related with the true factor matrices as
\begin{align}
    \left\{\begin{array}{l}
        \bm{\hat{A}}_1 = \bm{{A}}_1\bm{\Lambda}_1\bm{\Pi}_1 + \bm{E}_1\\
        \bm{\hat{B}}_1 = \bm{{B}}_1\bm{\Lambda}_2\bm{\Pi}_1 + \bm{E}_2\\
        \bm{\hat{C}}_1 = \bm{{C}}\bm{\Lambda}_3\bm{\Pi}_1 + \bm{E}_3
      \end{array}\right.
\end{align}
and
\begin{align}
    \left\{\begin{array}{l}
        \bm{\hat{A}}_2 = \bm{{A}}_2\bm{\Gamma }_1\bm{\Pi}_2 + \bm{\tilde{E}}_1\\
        \bm{\hat{B}}_2 = \bm{{B}}_2\bm{\Gamma }_2\bm{\Pi}_2 + \bm{\tilde{E}}_2\\
        \bm{\hat{C}}_2 = \bm{{C}}\bm{\Gamma }_3\bm{\Pi}_2 + \bm{\tilde{E}}_3
      \end{array}\right.
\end{align}
where $\{\bm{\Lambda}_1,\bm{\Lambda}_2,\bm{\Lambda}_3,\}$ and
$\{\bm{\Gamma}_1,\bm{\Gamma}_2,\bm{\Gamma}_3,\}$ are unknown
nonsingular diagonal matrices satisfying
$\bm{\Lambda}_1\bm{\Lambda}_2\bm{\Lambda}_3=\bm{I}$ and
$\bm{\Gamma}_1\bm{\Gamma}_2\bm{\Gamma}_3=\bm{I}$, $\{\bm{\Pi}_i\}$
are unknown permutation matrices, $\{\bm{E}_1,\bm{E}_2,\bm{E}_3\}$
and $\{\bm{\tilde{E}}_1,\bm{\tilde{E}}_2,\bm{\tilde{E}}_3\}$ are
estimation errors.

\subsection{The Scaling Ambiguity Issue}
First, we show that, when only a single phase is considered, why
the DOA estimation is infeasible in the scenario of
$\text{rank}(\boldsymbol{G})=1$. Recall that
$\bm{b}_i(\theta)=\bm{G}^T\bm{\Phi}_{i}\bm{a}(\theta)$. Hence, we
can write $\bm{B}_i=\bm{G}^T\bm{\Phi}_i\bm{\varXi}$, where
$\bm{\varXi}\triangleq[\bm{a}(\theta_1)\cdots
\bm{a}(\theta_K)]\in\mathbb{C}^{N\times K}$. If the rank of the
channel matrix $\bm{G}$ is greater than one, then we can employ a
correlation-based method \cite{7914672} to extract the parameter
$\theta_k$ from each column of $\bm{\hat{B}}_1$, i.e.,
\begin{align}
    \hat{\theta}_{k}=\mathop{\arg\max}\limits_{\theta_k}\frac{|\bm{\hat{b}}_{1,k}^{H}\bm{b}_1(\theta_k)|}
        {\Vert\bm{\hat{b}}_{1,k}\Vert_2\Vert\bm{b}_1(\theta_k)\Vert_2}
        \label{corr}
\end{align}
where $\bm{\hat{b}}_{1,k}$ is the $k$th column of $\bm{\hat{B}}_1$.
In this case, only using the
received signals from a single phase suffices to recover
parameters of interest. Nevertheless, this method fails when
$\text{rank}(\bm{G})=1$. The reason can be explained as follows.
Specifically, when $\text{rank}(\bm{G})=1$, $\bm{G}$ can be
expressed as $\bm{G}=\sigma\bm{u}\bm{v}^T$. In this case, we have
$\bm{\hat{B}}_1=
\sigma\bm{v}\bm{u}^T\bm{\Phi}_1\bm{\varXi}\bm{\Lambda}_2$, where,
for simplicity, the unknown permutation matrix and the estimation
error are neglected. The $k$th column of $\bm{\hat{B}}_1$ is thus
given by
\begin{align}
\bm{\hat{b}}_{1,k}&=
\sigma\bm{v}\bm{u}^T\bm{\Phi}_1\bm{a}(\theta_k)[\bm{\Lambda
}_2]_{k,k}
\end{align}
where $[\bm{\Lambda }_2]_{k,k}$ denotes the $k$th diagonal element
of $\bm{\Lambda }_2$. Here
$\bm{u}^T\bm{\Phi}_1\bm{a}(\theta_k)[\bm{\Lambda }_2]_{k,k}$ is a
complex scalar. Due to the coupling between the unknown scalar
$\bm{u}^T\bm{\Phi}_1\bm{a}(\theta_k)$ and the unknown scalar
$[\bm{\Lambda }_2]_{k,k}$, the parameter $\theta_k$ cannot be
uniquely identified.

From the above discussion, we see that when
$\text{rank}(\bm{G})=1$, using the factor matrices
$\{\bm{\hat{A}}_1,\bm{\hat{B}}_1,\bm{\hat{C}}_1\}$ obtained from
the first phase alone cannot uniquely identify the targets'
parameters. In the following, we will show how to utilize the
estimated factor matrices from two different phases to resolve
this ambiguity.

\subsection{Column Alignment for Factor Matrices}
To leverage the estimated factor matrices, we first
remove the permutation ambiguity between phase $1$ and phase $2$.
Notice that
both $\bm{\hat{C}}_1$ and $\bm{\hat{C}}_2$ are associated with a
common matrix $\boldsymbol{C}$. This fact can be used to remove
the relative permutation between $\bm{\hat{C}}_1$ and
$\bm{\hat{C}}_2$. Define
\begin{align}
   \rho_{k_1,k_2}=\frac{|(\bm{\hat{c}}_{1,k_1})^{H}\bm{\hat{c}}_{2,k_2}|}
   {\Vert \bm{\hat{c}}_{1,k_1}\Vert_2 \Vert \bm{\hat{c}}_{2,k_2}\Vert_2}
\end{align}
where $\bm{\hat{c}}_{1,k_1}$ and $\bm{\hat{c}}_{2,k_2}$ are,
respectively, the $k_1$th and $k_2$th column of $\bm{\hat{C}}_1$
and $\bm{\hat{C}}_2$. Since $\boldsymbol{C}$ has distinctive
columns, with each column characterized by a different time
delay parameter, $\rho_{k_1,k_2}$ achieves the largest value when
$\bm{\hat{c}}_{1,k_1}$ and $\bm{\hat{c}}_{2,k_2}$ correspond to
the same target. Define a permutation matrix
$\bm{{\Pi}}_3\triangleq[\bm{e}_{\pi(1)}\phantom{0}\cdots\phantom{0}
\bm{e}_{\pi(K)}]^T\in{\{0,1\}}^{K\times K}$, where
$\bm{e}_{\pi(k)}$ is a standard basis vector, and
$\pi(k)=\arg\max_{k_2}\{\rho_{k,k_2}\}_{k_2=1}^{K}$. Ignoring
estimation errors, we should have
\begin{align}
    \bm{\Pi}_2= \bm{\Pi}_1\bm{{\Pi}}_3
\end{align}
Then we can utilize $\bm{{\Pi}}_3$ to remove the permutation
between $\{\bm{\hat{A}}_1\}$ and $\{\bm{\hat{A}}_2\}$,
$\{\bm{\hat{B}}_1\}$ and $\{\bm{\hat{B}}_2\}$,
$\{\bm{\hat{C}}_1\}$ and $\{\bm{\hat{C}}_2\}$. Specifically,
defining $\bm{\tilde{A}}_1\triangleq\bm{\hat{A}}_1\bm{{\Pi}}_3$,
$\bm{\tilde{B}}_1\triangleq\bm{\hat{B}}_1\bm{{\Pi}}_3$,
$\bm{\tilde{C}}_1\triangleq\bm{\hat{C}}_1\bm{{\Pi}}_3$, we have
\begin{align}
\bm{\tilde{A}}_1&=\bm{A}_1\bm{\Lambda}_1\bm{\Pi}_1\bm{{\Pi}}_3 +
\bm{E}_1\bm{{\Pi}}_3=\bm{A}_1\bm{\Lambda}_1\bm{\Pi}_2 + \bm{E}_1\bm{{\Pi}}_3\\
\bm{\tilde{B}}_1&=\bm{B}_1\bm{\Lambda}_2\bm{\Pi}_1\bm{{\Pi}}_3 +
\bm{E}_2\bm{{\Pi}}_3=\bm{B}_1\bm{\Lambda}_2\bm{\Pi}_2 + \bm{E}_2\bm{{\Pi}}_3\\
\bm{\tilde{C}}_1&=\bm{C}\bm{\Lambda}_3\bm{\Pi}_1\bm{{\Pi}}_3 +
\bm{E}_3\bm{{\Pi}}_3=\bm{C}\bm{\Lambda}_3\bm{\Pi}_2 + \bm{E}_3\bm{{\Pi}}_3
\end{align}
Ignoring the permutation matrix $\bm{\Pi}_2$, we obtain
\begin{align}
    \left\{\begin{array}{l}
        \bm{\tilde{A}}_1 =  \bm{{A}}_1\bm{\Lambda }_1+\bm{E}_1\bm{{\Pi}}_3\\
        \bm{\hat{A}}_2 =  \bm{{A}}_2\bm{\Gamma }_1 + \bm{\tilde{E}}_1\\
      \end{array}\right.
\label{Aeq}
\end{align}
\begin{align}
    \left\{\begin{array}{l}
        \bm{\tilde{B}}_1 =\bm{B}_1\bm{\Lambda}_2 + \bm{E}_2\bm{{\Pi}}_3\\
        \bm{\hat{B}}_2 =\bm{{B}}_2\bm{\Gamma }_2 + \bm{\tilde{E}}_2\\
      \end{array}\right.
\label{Beq}
\end{align}
and
\begin{align}
    \left\{\begin{array}{l}
        \bm{\tilde{C}}_1 =\bm{C}\bm{\Lambda}_3 + \bm{E}_3\bm{{\Pi}}_3\\
        \bm{\hat{C}}_2 =\bm{{C}}\bm{\Gamma }_3 + \bm{\tilde{E}}_3\\
      \end{array}\right.
\label{Ceq}
\end{align}
Now we have column-aligned $\bm{\tilde{A}}_1$ and
$\bm{\hat{A}}_2$, $\bm{\tilde{B}}_1$ and $\bm{\hat{B}}_2$,
$\bm{\tilde{C}}_1$ and $\bm{\hat{C}}_2$, i.e., the same columns of
each pair of two factor matrices are associated with the same
target. Note that both $\bm{\tilde{C}}_1$ and $\bm{\hat{C}}_2$ are
estimated as a Vandermonde matrix based on the estimated
generators. Hence theoretically we should have $\bm{\Gamma
}_3\bm{\Lambda }_{3}^{-1}=\boldsymbol{I}$.

\subsection{Joint DOA, Time Delay and Doppler Estimation}
Based on the column-aligned factor matrices from two phases, we
discuss how to jointly estimate DOA, time delay and Doppler shift.
Since $\bm{B}_i=\bm{G}^T\bm{\Phi}_i\bm{\varXi}$, and
$\bm{G}=\sigma\bm{u}\bm{v}^T$, (\ref{Beq}) can be rewritten as
\begin{align}
    \left\{\begin{array}{l}
\bm{\tilde{B}}_1 =  \sigma\bm{v}\bm{u}^T\bm{\Phi}_1\bm{\varXi}\bm{\Lambda }_2+\bm{E}_2\bm{{\Pi}}_3\\
\bm{\hat{B}}_2 =
\sigma\bm{v}\bm{u}^T\bm{\Phi}_2\bm{\varXi}\bm{\Gamma
}_2+\bm{\tilde{E}}_2
      \end{array}\right.
\label{BeqSvd}
\end{align}
Also, (\ref{Aeq}) can be rewritten as
\begin{align}
    \left\{\begin{array}{l}
\bm{\tilde{A}}_1 =  \sigma\bm{W}^T\bm{v}\bm{u}^T\bm{\Phi}_1\bm{\varXi}\bm{\Lambda }_1
\circledast \bm{D}+\bm{E}_1\bm{{\Pi}}_3\\
\bm{\hat{A}}_2 =
\sigma\bm{W}^T\bm{v}\bm{u}^T\bm{\Phi}_2\bm{\varXi}\bm{\Gamma
}_1\circledast \bm{D}+\bm{\tilde{E}}_1
      \end{array}\right.
\label{AeqSvd}
\end{align}
where $\bm{D}\triangleq[\bm{d}(\nu_1)\phantom{0}\cdots\phantom{0}
\bm{d}(\nu_K)]\in\mathbb{C}^{P\times K}$. Note both $\{\bm{A}_i\}$
and $\{\bm{B}_i\}$ contain the DOA information. To harness the IRS
illumination diversity across two phases, we define a new vector
$\bm{\hat{r}}_k^B\in\mathbb{C}^M$, in which the $m$th entry is
calculated by the element-wise division of
$[\bm{\tilde{B}}_1]_{m,k}$ and $[\bm{\hat{B}}_2]_{m,k}$, i.e.,
\begin{align}
[\bm{\hat{r}}^B_k]_m &\triangleq\frac{[\bm{\tilde{B}}_1]_{m,k}}{
[\bm{\hat{B}}_2]_{m,k}} \nonumber \\
&=\frac{\sigma[\bm{v}]_m\bm{u}^T\bm{\Phi}_1\bm{a}(\theta_k)[\bm{\Lambda
}_2]_{k,k}}{\sigma[\bm{v}]_m\bm{u}^T\bm{\Phi}_2\bm{a}(\theta_k)[\bm{\Gamma
}_2]_{k,k}} + \epsilon_{m,k}\nonumber\\
&=\frac{\bm{u}^T\bm{\Phi}_1\bm{a}(\theta_k)[\bm{\Lambda
}_2]_{k,k}}{\bm{u}^T\bm{\Phi}_2\bm{a}(\theta_k)[\bm{\Gamma
}_2]_{k,k}} + \epsilon_{m,k}
\label{rA}
\end{align}
Similarly, define $\bm{\hat{r}}^A_k\in\mathbb{C}^P$, we have
\begin{align}
[\bm{\hat{r}}^A_k]_p&\triangleq\frac{[\bm{\tilde{A}}_1]_{p,k}}{
[\bm{\hat{A}}_2]_{p,k}} \nonumber \\
&=\frac{\sigma\bm{w}^T_p\bm{v}
\bm{u}^T\bm{\Phi}_1\bm{a}(\theta_k)[\bm{\Lambda
}_1]_{k,k}}{\sigma\bm{w}^T_p\bm{v}\bm{u}^T\bm{\Phi}_2\bm{a}(\theta_k)[\bm{\Gamma
}_1]_{k,k}} +\varepsilon_{p,k} \nonumber \\
&=\frac{\bm{u}^T\bm{\Phi}_1\bm{a}(\theta_k)[\bm{\Lambda
}_1]_{k,k}}{\bm{u}^T\bm{\Phi}_2\bm{a}(\theta_k)[\bm{\Gamma
}_1]_{k,k}} +\varepsilon_{p,k}
\label{rB}
\end{align}
where $\bm{w}_p$ denotes the $p$th column of $\bm{W}$, both
$\epsilon_{m,k}$ and $\varepsilon_{p,k}$ are noise terms. We now
discuss how to recover the DOA parameter from (\ref{rA}) and
(\ref{rB}). Define
\begin{align}
    \hat{\gamma}(\theta_k)\triangleq \frac{1}{M}
    \sum_{m=1}^{M}[\bm{\hat{r}}^B_k]_m \cdot \frac{1}{P}\sum_{p=1}^{P}[\bm{\hat{r}}^A_k]_p
\end{align}
Ignoring the noise term, $\hat{\gamma}(\theta_k)$ is equivalent to
\begin{align}
\hat{\gamma}(\theta_k)&=\frac{\bm{u}^T\bm{\Phi}_1\bm{a}(\theta_k)
[\bm{\Lambda
}_2]_{k,k}}{\bm{u}^T\bm{\Phi}_2\bm{a}(\theta_k)[\bm{\Gamma
}_2]_{k,k}} \cdot \frac{\bm{u}^T\bm{\Phi}_1\bm{a}(\theta_k)
[\bm{\Lambda }_1]_{k,k}}{\bm{u}^T\bm{\Phi}_2\bm{a}(\theta_k)[\bm{\Gamma }_1]_{k,k}} \nonumber\\
   &=\left(\frac{\bm{u}^T\bm{\Phi}_1\bm{a}(\theta_k)}
   {\bm{u}^T\bm{\Phi}_2\bm{a}(\theta_k)}\right)^2[\bm{\Lambda}_2]_{k,k}
   [\bm{\Gamma }_2]_{k,k}^{-1}[\bm{\Lambda }_1]_{k,k}[\bm{\Gamma }_1]_{k,k}^{-1} \nonumber\\
   &=\left(\frac{\bm{u}^T\bm{\Phi}_1\bm{a}(\theta_k)}{\bm{u}^T
   \bm{\Phi}_2\bm{a}(\theta_k)}\right)^2[\bm{\Gamma }_3]_{k,k}[\bm{\Lambda }_3]_{k,k}^{-1} \nonumber\\
   &\overset{(a)}=\left(\frac{\bm{u}^T\bm{\Phi}_1\bm{a}(\theta_k)}
   {\bm{u}^T\bm{\Phi}_2\bm{a}(\theta_k)}\right)^2
   \label{unamTheta}
\end{align}
where $(a)$ comes from the fact that
$\bm{\Lambda}_1\bm{\Lambda}_2\bm{\Lambda}_3=\bm{I}$,
$\bm{\Gamma}_1\bm{\Gamma}_2\bm{\Gamma}_3=\bm{I}$, and $[\bm{\Gamma
}_3]_{k,k}[\bm{\Lambda }_3]_{k,k}^{-1}=1$. As for now, we see that
the scaling ambiguities in (\ref{rA}) and (\ref{rB}) are
effectively removed and an unambiguous estimate of $\theta$ can be
obtained.

Based on the above relationship (\ref{unamTheta}), we can estimate
the target's DOA via the following criterion
\begin{equation}
    \begin{aligned}
\hat{\theta}_k=\mathop{\arg\min}\limits_{\theta}
\quad&\Vert{{\hat{\gamma}}}(\theta_k)- {\gamma}(\theta)\Vert^2_2\\
    \mathrm{s.t.}\quad&{\gamma}(\theta)=
    \left(\frac{\bm{u}^T\bm{\Phi}_1\bm{a}(\theta)}{\bm{u}^T\bm{\Phi}_2\bm{a}(\theta)}\right)^2\\
    &\theta \in\mathcal{D}_{\theta}
\end{aligned}
\label{Opeq}
\end{equation}
where $\mathcal{D}_{\theta}$ is the feasible region of $\theta$
and the above problem can be easily solved by a one-dimensional
search.

Note that the $k$th column of $\bm{A}_i$ is characterized by both
$\theta_k$ and $\nu_k$. Specifically, the $k$th column of
$\bm{A}_i$ and the $k$th column of $\bm{B}_i$ are related as
\begin{align}
\boldsymbol{z}_i(\theta_k,\nu_k)=(\boldsymbol{W}^T\boldsymbol{b}_i(\theta_k))\circledast(\boldsymbol{d}(\nu_k))
\end{align}
After the DOA is estimated, define
$\bm{\check{B}}_i=\bm{G}^T\bm{\Phi}_i\bm{\hat{\varXi}}\in\mathbb{C}^{M\times
K}$ with $\bm{\hat{\varXi}}\triangleq[\bm{a}(\hat{\theta}_1)\cdots
\bm{a}(\hat{\theta}_K)]$, and define
$\bm{\check{A}}_i\in\mathbb{C}^{P\times K}$ with
$[\bm{\check{A}}_1]_{p,k}=[\bm{\tilde{A}}_1]_{p,k}/[\bm{W}^T\bm{\check{B}}_1]_{p,k}$,
$[\bm{\check{A}}_2]_{p,k}=[\bm{\hat{A}}_2]_{p,k}/[\bm{W}^T\bm{\check{B}}_2]_{p,k}$.
Note that each column of $\bm{\check{A}}_i$ is characterized by
the associated Doppler shift $\nu_k$. Hence, the Doppler shift
$\nu_k$ can be estimated via a correlation-based scheme
\cite{7914672} as
\begin{align}
\hat{\nu}_{i,k}=\mathop{\arg\max}\limits_{\nu_k}\frac{|\bm{\check{a}}_{i,k}^{H}\bm{d}(\nu_k)|}
        {\Vert\bm{\check{a}}_{i,k}\Vert_2\Vert\bm{d}(\nu_k)\Vert_2}
\end{align}
where $\bm{\check{a}}_{i,k}$ denotes the $k$th column of
$\bm{\check{A}}_i$. We then compute the average of the two
estimates as the final estimate of the Doppler shift, i.e.,
$\hat{\nu}_{k}=(\hat{\nu}_{1,k}+\hat{\nu}_{2,k})/2$. The velocity
estimate of the $k$th target can be calculated as
$\hat{v}_k=\hat{\nu}_k c/2f_c$. The round-trip time delay
$\{\hat{\tau}_{i,k}\}$ can be calculated from the estimated
generators $\{\hat{t}_{i,k}\}$ in (\ref{EstGen}) as
\begin{align}
    \hat{\tau}_{i,k} = \frac{\mathrm{arg}(\hat{t}_{i,k})}{-2\pi \Delta f}
\end{align}
where $\mathrm{arg}(\hat{t}_{i,k})$ denotes the argument of the complex number $\hat{t}_{i,k}$.
Similarly, we obtain
$\hat{\tau}_{k}=(\hat{\tau}_{1,k}+\hat{\tau}_{2,k})/2$.

\section{CRB Analysis}
\label{CRBAnalysis}
In this section, we provide a CRB analysis of the estimation
problem considered in this paper. For the $P\times M\times L$
tensor observation $\bm{\mathcal{Y}}_i$ considered in
(\ref{CPform}), we have
\begin{align}
\bm{\mathcal{Y}}_i =
\sum_{k=1}^{K}\bm{z}_i(\theta_k,\nu_k)\circ\bm{b}_i(\theta_k)\circ\alpha_k\bm{f}(\tau_k)
+ \bm{\mathcal{N}}_i
\end{align}
We write the unknown target parameters as
\begin{align}
    \bm{\zeta}=[\bm{\theta}^T\phantom{0}\bm{\nu}^T\phantom{0}
    \bm{\tau}^T]\in\mathbb{R}^{1\times 3K}
\end{align}
where
$\bm{\theta}\triangleq[\theta_1\phantom{0}\cdots\phantom{0}\theta_K]^T$,
$\bm{\nu}\triangleq[\nu_1\phantom{0}\cdots\phantom{0}\nu_K]^T$ and
$\bm{\tau}\triangleq[\tau_1\phantom{0}\cdots\phantom{0}\tau_K]^T$.
The log-likelihood function of the parameter vector $\bm{\zeta}$
can be expressed as
\begin{align}
    \mathcal{L}(\bm{\zeta})
    &=\sum_{i=1}^{2}-\tilde{D}_i-\frac{1}{\sigma_i^2}
    \Vert \bm{\mathcal{Y}}_{i,(1)}-\bm{A}_i(\bm{C}\odot \bm{B}_i)^T\Vert_F^2 \nonumber \\
    &=\sum_{i=1}^{2}-\tilde{D}_i-\frac{1}{\sigma_i^2}\Vert
    \bm{\mathcal{Y}}_{i,(2)}-\bm{B}_i(\bm{C}\odot \bm{A}_i)^T\Vert_F^2 \nonumber \\
    &=\sum_{i=1}^{2}-\tilde{D}_i-\frac{1}{\sigma_i^2}\Vert
    \bm{\mathcal{Y}}_{i,(3)}-\bm{C}(\bm{B}_i\odot \bm{A}_i)^T\Vert_F^2
\end{align}
where $\tilde{D}_i\triangleq PML\ln(\pi\sigma_i^2)$, and
$\bm{\mathcal{Y}}_{i,(j)}$ denotes the mode-$j$ unfolding of
$\bm{\mathcal{Y}}_i$. Then the fisher information matrix (FIM) for
$\bm{\zeta}$ is given by
\begin{align}
    \bm{\Omega}(\bm{\zeta})= \mathbb{E}\left\{\left(\frac{\partial
    \mathcal{L}(\bm{\zeta})}{\partial \bm{\zeta}}\right)^H
    \left(\frac{\partial \mathcal{L}(\bm{\zeta})}{\partial \bm{\zeta}}\right)\right\}
 \end{align}
To calculate $\bm{\Omega}(\bm{\zeta})$, we first compute the
partial derivative of $\mathcal{L}(\bm{\zeta})$ with respect to
$\bm{\zeta}$ and then calculate its expectation.

\subsection{Partial Derivative of $\mathcal{L}(\bm{\zeta})$}
Following a similar procedure as in \cite{7914672,10274514}, the
partial derivative of $\mathcal{L}(\bm{\zeta})$ w.r.t. $\theta_k$
can be calculated as

\begin{align}
    \frac{\partial \mathcal{L}(\bm{\zeta})}{\partial
    \theta_k}
    =&\sum_{i=1}^{2}\frac{2}{\sigma_i^2}\Re
    \bigg\{\bm{e}_k^T\left(\bm{{C}}\odot\bm{{B}}_i\right)^T \nonumber\\
    {}&\times \left(\bm{\mathcal{Y}}_{i,(1)}-\bm{A}_i
    \left(\bm{{C}}\odot\bm{{B}}_i\right)^T
    \right)^H\bm{A}_{i,\theta}^{\prime}\bm{e}_k\bigg\}\nonumber\\
    +&{}\sum_{i=1}^{2}\frac{2}{\sigma_i^2}\Re \bigg\{
    \bm{e}_k^T(\bm{{C}}\odot\bm{{A}}_i)^T  \nonumber\\
    {}&\times\left(\bm{\mathcal{Y}}_{i,(2)}-\bm{B}_i
    \left(\bm{{C}}\odot\bm{{A}}_i\right)^T
            \right)^{H}\bm{B}_{i,\theta}^{\prime}\bm{e}_k \bigg\}
\end{align}
in which
$\bm{A}_{i,\theta}^{\prime}\triangleq[\bm{{a}}_{i,\theta_1}^{\prime}\phantom{0}
\cdots\phantom{0}\bm{{a}}_{i,\theta_K}^{\prime}]\in\mathbb{C}^{P\times
K}$ and
$\bm{B}_{i,\theta}^{\prime}\triangleq[\bm{{b}}_{i,\theta_1}^{\prime}\phantom{0}
\cdots\phantom{0}\bm{{b}}_{i,\theta_K}^{\prime}]\in\mathbb{C}^{M\times
K}$, with $\bm{{a}}_{i,\theta_k}^{\prime
}\triangleq\frac{\partial[\bm{{A}}_i]_{:,k}}{\partial\theta_k}$
and $\bm{{b}}_{i,\theta_k}^{\prime
}\triangleq\frac{\partial[\bm{{B}}_i]_{:,k}}{\partial\theta_k}$.
$\bm{e}_k$ is a standard basis vector with $k$ as the index of its
nonzero element. Similarly, the partial derivatives w.r.t. other
parameters can be calculated. The details are omitted here for
brevity.

\subsection{Calculation of FIM $\bm{\Omega}(\bm{\zeta})$}
To calculate the FIM $\bm{\Omega}(\bm{\zeta})$, we first calculate
the entries in the diagonal blocks. Define $u=K(k_1-1)+k_1$ and
$v=K(k_2-1)+k_2$. The $(k_1,k_2)$th element in the block related
to $\bm{\theta}$, can be calculated as
\begin{multline}
\mathbb{E}\left\{ \left(\frac{\partial
\mathcal{L}(\bm{\zeta})}{\partial
\theta_{k_1}}\right)^*\left(\frac{\partial
\mathcal{L}(\bm{\zeta})}{\partial
\theta_{k_2}}\right)\right\}\\
=2\Re\{ [\bm{C}_{\bm{n}_{A,\theta}}]_{u,v}\}
+2\Re\{ [\bm{C}_{\bm{n}_{A_{\theta},B}}]_{u,v}\}\\
+2\Re\{ [\bm{C}_{\bm{n}_{B,A_{\theta}}}]_{u,v}\}
+2\Re\{ [\bm{C}_{\bm{n}_{B,\theta}}]_{u,v}\}
\end{multline}
where
\begin{align}
\bm{C}_{\bm{n}_{A_{\theta},B}}=&\sum_{i=1}^{2}\frac{1}{\sigma_i^4}
\left(\bm{A}_{i,\theta}^{\prime}\otimes(\bm{C}\odot\bm{B}_i)\right)^T\bm{C}_{1,2}^i \nonumber \\
&\times \left(\bm{B}_{i,\theta}^{\prime}\otimes (\bm{C}\odot\bm{A}_i)\right)^*\\
\bm{C}_{\bm{n}_{B,A_{\theta}}}=&\sum_{i=1}^{2}\frac{1}{\sigma_i^4}
\left(\bm{B}_{i,\theta}^{\prime}\otimes(\bm{C}\odot\bm{A}_i)\right)^T\bm{C}_{2,1}^i \nonumber \\
&\times  \left(\bm{A}_{i,\theta}^{\prime}\otimes(\bm{C}\odot\bm{B}_i)\right)^*\\
\bm{C}_{\bm{n}_{B,\theta}}=&\sum_{i=1}^{2}\frac{1}{\sigma_i^2}
\left(\bm{B}_{i,\theta}^{\prime}\otimes(\bm{C}\odot\bm{A}_i)\right)^T \nonumber \\
&\times\left(\bm{B}_{i,\theta}^{\prime}\otimes(\bm{C}\odot\bm{A}_i)\right)^*
\end{align}
and
\begin{align}
\bm{C}_{j_1,j_2}^i\triangleq\mathbb{E}\{\mathrm{vec}(\bm{\mathcal{N}}_{i,(j_1)}^H)
\mathrm{vec}(\bm{\mathcal{N}}_{i,(j_2)}^H)^H\}
\label{CovNoise}
\end{align}
in which $i$ is phase index and the details will be discussed
later. Similarly, the entries in other blocks can be derived The details are omitted due to space limit.

Now we compute $\bm{C}_{j_1,j_2}^i$ defined in (\ref{CovNoise}).
Note the entries in $\bm{\mathcal{N}}_i$ are all i.i.d Gaussian
random variables, we have
\begin{equation}
    \mathbb{E}\{n_{i,p_1,m_1,l_1}n_{i,p_2,m_2,l_2}^*\}=\begin{cases}
        \sigma^2_i,&p_1=p_2,m_1=m_2,l_1=l_2\\
    0,&\text{otherwise}
    \end{cases}
\end{equation}
Based on the arrangements of elements of $\bm{\mathcal{N}}_i$
under different unfolding modes, the $PML$ nonzero entries in
$\bm{C}_{j_1,j_2}^i\in\mathbb{C}^{PML\times PML}$ can be given as
\begin{equation}
    [\bm{C}_{j_1,j_2}^i]_{u,v}=\begin{cases}
        \sigma^2_i,&u=\hbar_{j_1}(p,m,l),v=\hbar_{j_2}(p,m,l)\\
    0,&\text{otherwise}
    \end{cases}
\end{equation}
where $ \forall j_1\neq j_2, 1\leq j_1\leq3, 1\leq j_2\leq3,$ and
\begin{align}
    \hbar_1(p,m,l)&= m+(l-1)M+(p-1)ML\\
    \hbar_2(p,m,l)&= p+(l-1)P+(m-1)PL\\
    \hbar_3(p,m,l)&= p+(m-1)P+(l-1)PM
\end{align}
After obtaining the FIM $\bm{\Omega}$, the CRB can be calculated as
\begin{align}
    \text{CRB}(\bm{\zeta})=\bm{\Omega}^{-1}(\bm{\zeta})
\end{align}

\begin{figure}[htbp]
    \centering
    {\includegraphics[width=.9\linewidth]{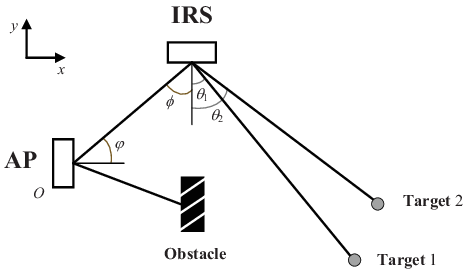}}
    \caption{Simulation setup (top view).}
    \label{simulation_setup}
 \end{figure}

\section{Simulation Results}
\label{Simulations} We now present numerical results to evaluate
the estimation performance of the proposed method for NLOS target
sensing. We examine a two-dimensional scenario, as illustrated in
Fig. \ref{simulation_setup}, where the AP and the IRS are located
at coordinates $\bm{p}_{\text{AP}}=[0,0]^T$ and
$\bm{p}_{\text{IRS}}=[100,100]^T$m, respectively. In our
simulations, the system carrier frequency is set to $f_c=60$ GHz,
and the distance between two adjacent antenna elements $d$ is set
to half of the signal wavelength. The number of antennas at the AP
and the number of reflecting elements at the IRS are set to $M=16$
and $N=32$, respectively. The length of an effective OFDM symbol
is set to $T_{\text{d}}=2\mu\text{s}$ and the length of the CP is
set to $T_{\text{cp}}=1\mu\text{s}$ \cite{9420261}. The pulse
repetition interval is set to $T_{\text{PRI}}=8 \mu\text{s}$. The
channel of the AP-IRS link is generated based on the geometric
channel model and includes only a LOS path, i.e.
\begin{equation}
    \setlength{\abovedisplayskip}{5pt}
    \setlength{\belowdisplayskip}{5pt}
\begin{aligned}
\bm{G}=\varrho  \bm{a}_{\text{IRS}}(\phi
)\bm{a}_{\text{AP}}^H(\varphi)
\end{aligned}
\end{equation}
where $\varrho$ denotes the path loss between the AP and the IRS,
$\phi$ and $\varphi$ denote the angle of arrival (AOA) and angle
of departure (AOD), respectively.
The distance-dependent path loss
$\gamma$ follows a complex normal distribution
$\mathcal{CN}(0,10^{-0.1\kappa})$, where
$\kappa = a+10b\log_{10}(D)+\xi$.
Here, $\xi\sim\mathcal{N}(0,\sigma_{\xi}^2)$, and $D$ represents the
distance between the the AP and the IRS. The parameters $a$, $b$,
and $\sigma_{\xi}$ are set to $a=68$, $b=2$, and
$\sigma_{\xi}=5.8$ dB, as suggested in \cite{6834753,8386686}. In our
experiments, we consider $K=2$ targets, both located within the
angular range of $[30^\circ,45^\circ]$ with respect to the IRS.
The coordinates of the targets are set as
$\bm{p}_{1}=[533,-170]^T$m and $\bm{p}_{2}=[541,-245]^T$m. The
targets' radial velocities with respect to the IRS are set to
$v_1=16.66$ m/s and $v_2=-22$ m/s, respectively. The direct link
between the AP and the targets are blocked by obstacles. So the AP
has to detect these two targets via the IRS-assisted reflected
path. In our simulations, the targets' radar cross section (RCS)
is set to one \cite{10138058}.

In our experiments, the beamforming vector $\bm{w}_p$ is designed
to align its beam direction towards the IRS to maximize the
received signal power at the IRS. As discussed earlier, our
proposed method relies on leveraging the diversity in the
illumination pattern of the IRS across two different phases.
Suppose the target's DOA is within an interval that is assumed
\emph{a priori} known, i.e.
$\theta_k\in[\theta^{\mathrm{lb}},\theta^{\mathrm{ub}}]$, where
$\theta^{\mathrm{ub}}>\theta^{\mathrm{lb}}$. To design the IRS
coefficients, we partition the IRS into four subarrays, with each
subarray steering a beam towards an individual direction. The four
directions are devised such that the beam pattern covers the
spatial area specified by
$[\theta^{\mathrm{lb}},\theta^{\mathrm{ub}}]$. In different
phases, we can let each beam steer towards a different direction
to ensure that the radiating patterns in two phases are different.
Unless otherwise stated, the numbers of subcarriers and pulses, as
well as the transmit power, are set to $L=10$, $P=10$, and
$P_t=30$ dBm, respectively. The received signal-to-noise ratio
(SNR) is defined as
\begin{equation}
\begin{aligned}
    \text{SNR}\triangleq\frac{\| \bm{\mathcal{Y}}-\bm{\mathcal{N}}\|_F^2 }{\|\bm{\mathcal{N}} \|_F^2 }
\end{aligned}
\end{equation}
where $\bm{\mathcal{Y}}$ and $\bm{\mathcal{N}}$ represent the
received signal and the additive noise in (\ref{CPform}),
respectively. All results are averaged over $10^3$ Monte Carlo
runs.

\subsection{Performance Evaluation of The Proposed Method}
\begin{figure*}[htbp]
    \centering
    \begin{minipage}[t]{0.49\textwidth}
        \centering
        \subfloat[MSE and CRB of $\theta$.]{\includegraphics[width=.75\linewidth]{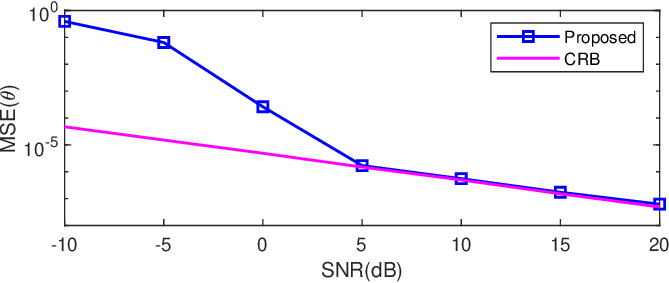}}\\
        \subfloat[MSE and CRB of $\nu$.]{\includegraphics[width=.75\linewidth]{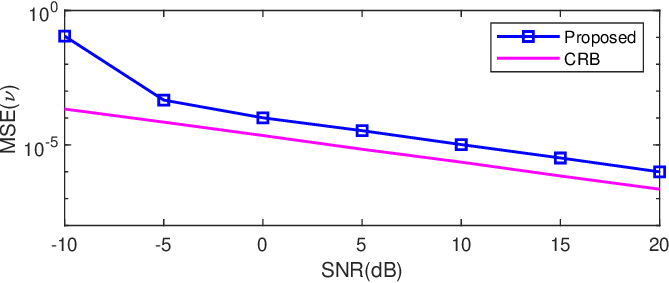}}\\
        \subfloat[MSE and CRB of $\tau$.]{\includegraphics[width=.75\linewidth]{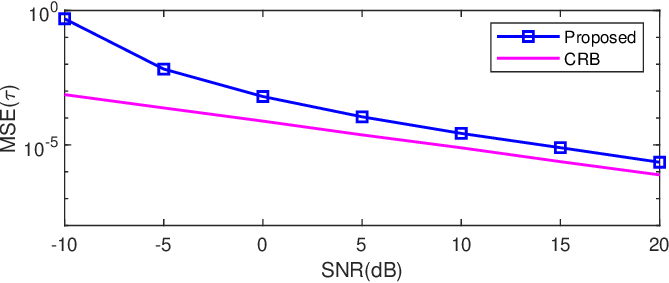}}
        \caption{MSEs and CRBs versus SNR.}
        \label{Mse_vs_Snr}
    \end{minipage}
    \begin{minipage}[t]{0.49\textwidth}
        \centering
    \subfloat[MSE and CRB of $\theta$.]{\includegraphics[width=.75\linewidth]{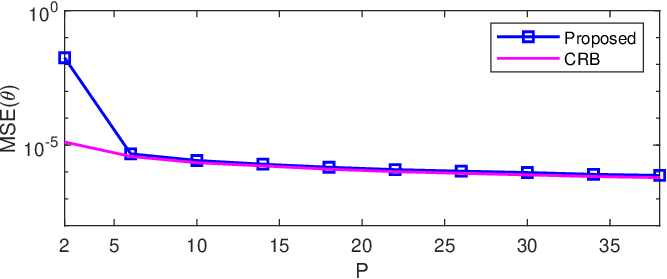}}\\
    \subfloat[MSE and CRB of $\nu$.]{\includegraphics[width=.75\linewidth]{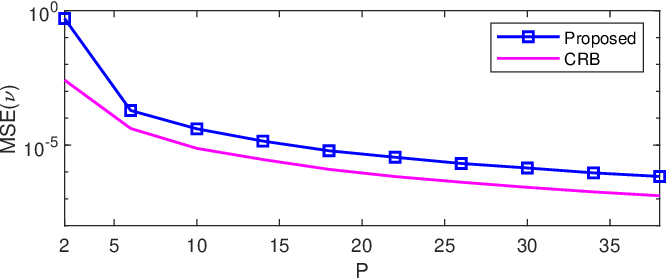}}\\
    \subfloat[MSE and CRB of $\tau$.]{\includegraphics[width=.75\linewidth]{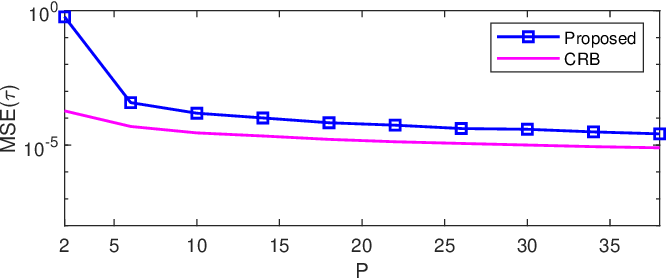}}
    \caption{MSEs and CRBs versus the number of pulses.}
    \label{Mse_vs_pulse}
    \end{minipage}
\end{figure*}

We first examine the performance of our proposed method in
estimating the target's parameters $\{\theta_k,\tau_k,\nu_k\}$.
The performance is evaluated by the mean square error (MSE), which
is defined as
\begin{equation}
    \setlength{\abovedisplayskip}{3pt}
    \setlength{\belowdisplayskip}{3pt}
\begin{aligned}
\text{MSE}(\zeta) = \frac{1}{K}\sum_{k=1}^{K}\mathbb{E} \left(\|
\zeta_k-\hat{\zeta}_k\|_2^2 \right)
\end{aligned}
\end{equation}
where $\hat{\zeta}$ denotes an estimate of the parameter $\zeta$,
which corresponds to one of the parameters $\{\theta,\tau,\nu\}$.
The MSE of our proposed method as a function of the SNR is plotted
in Fig. \ref{Mse_vs_Snr}(a)-\ref{Mse_vs_Snr}(c). The CRB results
for different sets of parameters are also included for comparison.
From Fig. \ref{Mse_vs_Snr}, we see that as the SNR increases, our
proposed method achieves an estimation accuracy that is close to
the theoretical lower bound. This result validates the efficiency
of the proposed method for NLOS target sensing. Specifically, the
MSE of the DOA is able to approach its CRB. The estimate of the
other two parameters (Doppler shift and delay) cannot exactly
attain their respective CRBs, which is probably because the
observation time/signal bandwidth is not long/large enough to
estimate these two parameters. Additionally, from Fig.
\ref{Mse_vs_Snr}, it is seen that the proposed method provides
accurate estimates of the target's parameters even in a relatively
low SNR regime, say $\text{SNR}=-5$dB. Notably, for NLOS sensing
tasks, the SNR is usually low due to the round-trip path loss and
reflection loss. Hence the ability of extracting parameters
reliably under a low SNR environment has a significant implication
in practice.

\begin{figure*}[htbp]
    \centering
    \begin{minipage}[t]{0.49\textwidth}
        \centering
        \subfloat[MSE and CRB of $\theta$.]{\includegraphics[width=.75\linewidth]{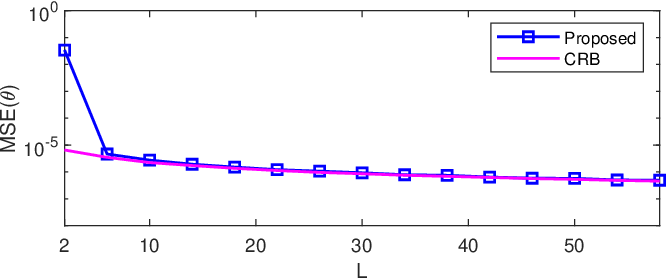}}\\
        \subfloat[MSE and CRB of $\nu$.]{\includegraphics[width=.75\linewidth]{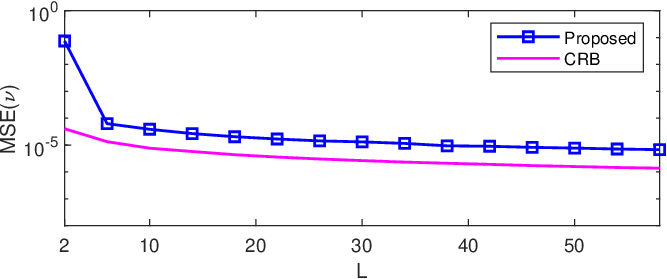}}\\
        \subfloat[MSE and CRB of $\tau$.]{\includegraphics[width=.75\linewidth]{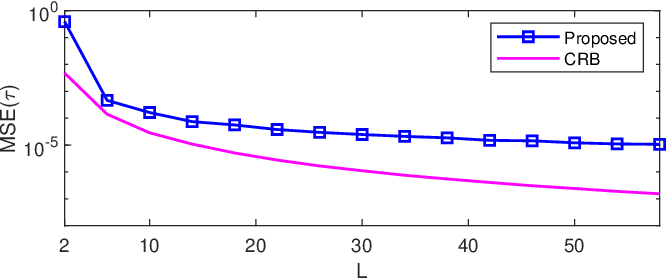}}
        \caption{MSEs and CRBs versus the number of subcarriers.}
        \label{Mse_vs_subcarrier}
    \end{minipage}
    \begin{minipage}[t]{0.49\textwidth}
        \centering
        \subfloat[MSE and CRB of $\theta$.]{\includegraphics[width=.75\linewidth]{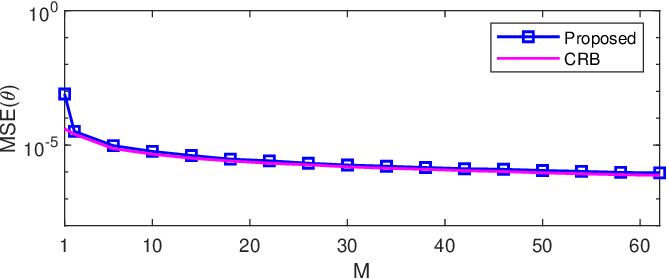}}\\
        \subfloat[MSE and CRB of $\nu$.]{\includegraphics[width=.75\linewidth]{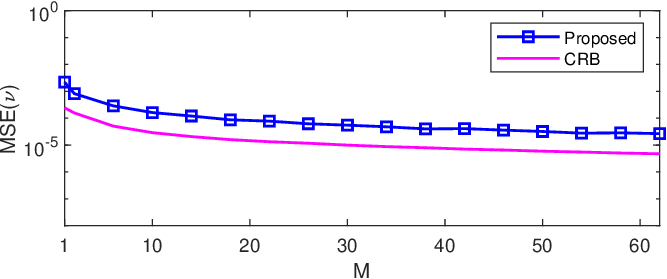}}\\
        \subfloat[MSE and CRB of $\tau$.]{\includegraphics[width=.75\linewidth]{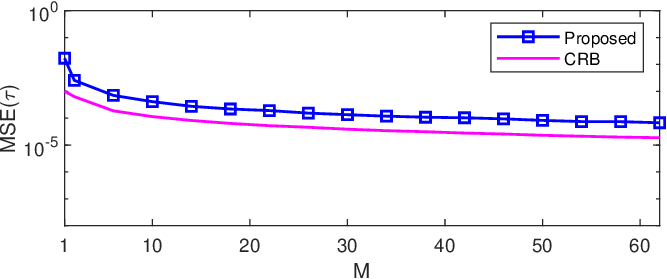}}
        \caption{MSEs and CRBs versus the number of AP's antennas.}
        \label{Mse_vs_antennas}
    \end{minipage}
\end{figure*}

Next, we plot the MSEs of the proposed method as a function of the
number of pulses $P$ in Fig.
\ref{Mse_vs_pulse}(a)-\ref{Mse_vs_pulse}(c), where the number of
subcarriers is set to $L=10$, and the SNR is set to $5$dB. We see
that the proposed method can achieve reliable sensing even with a
small number of pulses, for example, $P=5$, which corresponds to a
total sensing duration of $(2P+1)T_{\text{PRI}}=88\mu\text{s}$.
This result corroborates the efficiency of the proposed method for
NLOS sensing tasks.
In Fig. \ref{Mse_vs_subcarrier}(a)-\ref{Mse_vs_subcarrier}(c), we
depict the estimation performance of the proposed method as a
function of the number of subcarriers, where the number of pulses
is set to $P=10$ and the SNR is set to $5$ dB. The results show that
our proposed method can deliver accurate estimates of the target's
parameters even with a small number of subcarriers. We also
observe that our proposed method fails when the number of
subcarriers $L\leq 2$. This is because, when $P=10>K$ and
$M=16>K$, the uniqueness condition (\ref{eqn2}) is satisfied only
when $(L-1)\geq1$, implying that $L\geq 2$. Hence, the results
roughly coincide with our analysis concerning the uniqueness of the CP
decomposition.

In Fig. \ref{Mse_vs_antennas}(a)-\ref{Mse_vs_antennas}(c),
we plot the estimation performance of the proposed method versus
the number of AP's antennas. In this experiment, the number of
subcarriers and the number of pulses are set to $L=8$ and $P=8$,
respectively. The SNR is fixed at $5$dB. As expected, the
estimation performance improves with an increasing number of
antennas $M$. Furthermore, it is observed that the proposed method
can deliver decent performance even with a few number of
antennas employed at the AP. This result also corroborates well
with our analysis concerning the uniqueness condition of the CP
decomposition.

\subsection{Performance Comparison with The Existing Method \cite{10138058}}
To illustrate the superiority of the proposed method, we compare
it with the MLE-based method \cite{10138058}. For a fair
comparison, the AP-IRS channel is assumed to be Rician fading in
our simulations, i.e.
\begin{align}
    \bm{G}=\sqrt{\frac{\gamma}{1+\gamma}}
    \bm{G}^{\mathrm{LOS}}+\sqrt{\frac{1}{1+\gamma}}
    \bm{G}^{\mathrm{NLOS}}
\end{align}
where $\gamma$ is the Rician factor in dB while
$\bm{G}^{\mathrm{LOS}}$ and $\bm{G}^{\mathrm{NLOS}}$ are the LOS
and NLOS components, respectively. A typical value of the Rician
factor over the millimeter-wave (mmWave) band is $13$dB
\cite{6133581,7306533,9625066}, indicating that $\bm{G}$ is an
approximately rank-one matrix. In our experiments, we also
consider the cases where the Rician factor is set to $0$dB and
$5$dB, in order to more comprehensively examine the performance of
our proposed method under different channel conditions. Note that
\cite{10138058} employs a single-carrier signal to sense a single
static target. In contrast, this paper aims to sense multiple
moving targets based on the OFDM signal. To make a fair
comparison, we focus our simulations on a single static target.
The Doppler shift is set to $\nu =0$ in (\ref{eqn1}). It is
crucial to emphasize that our proposed method estimates not only
the DOA but also the Doppler shift and the time delay, whereas
\cite{10138058} can only estimate the DOA. In our experiments, we
assume there is one LOS path and four NLOS paths between the AP
and the IRS, resulting in $\mathrm{rank}(\bm{G})=5$. Also, for a
fair comparison, the number of measurements used for parameter
estimation is set the same for both methods.

\begin{figure}[htbp]
    \centering
    \begin{minipage}[t]{0.49\textwidth}
        \centering
    \subfloat[MSEs versus SNR, where $\gamma=0$dB.]
    {\includegraphics[width=.75\linewidth]{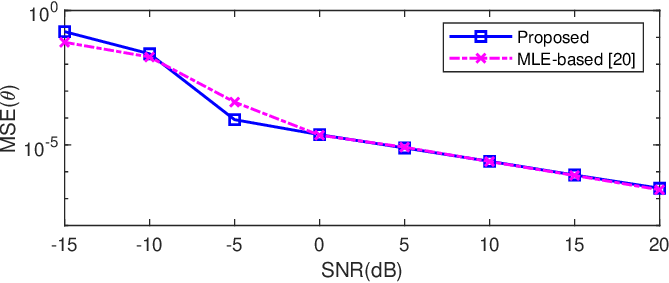}}\\
        \subfloat[MSEs versus SNR, where $\gamma=5$dB.]
        {\includegraphics[width=.75\linewidth]{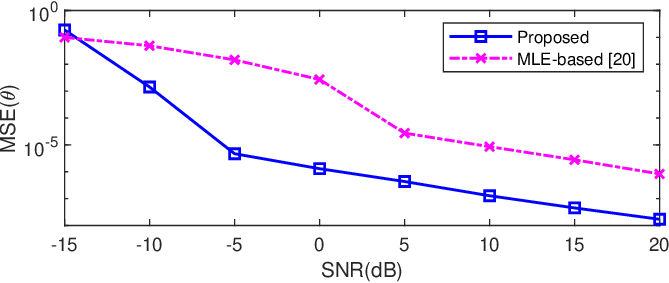}}\\
        \subfloat[MSEs versus SNR, where $\gamma=13$dB.]
        {\includegraphics[width=.75\linewidth]{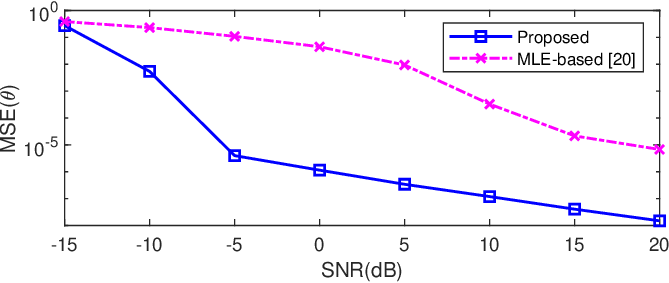}}
        \caption{MSEs achieved by respective algorithms versus SNR, where $K=1$ and $\mathrm{rank}(\bm{G})=5$.}
        \label{Comp_T1}
    \end{minipage}
\end{figure}

Fig.\ref{Comp_T1} depicts the MSE of the estimated DOA parameter
as a function of the SNR under different Rician factor values.
From Fig.\ref{Comp_T1}(b)-Fig.\ref{Comp_T1}(c), we observe that
our proposed algorithm presents a clear performance advantage over
the MLE method \cite{10138058}. This performance improvement
becomes more significant as the Rician factor increases. The
reason for this observation can be explained as follows. The work
\cite{10138058} requires additional degrees-of-freedom provided by
the AP-IRS channel in order to resolve the scaling ambiguity in
the DOA estimation. However, as the Rician factor increases, the
AP-IRS channel becomes an approximately rank-one matrix, yielding
an insufficient degrees-of-freedom for DOA estimation. As a
result, the method \cite{10138058} incurs a significant amount of
performance degradation as the Rician factor increases. In
contrast to \cite{10138058}, our proposed method removes the
scaling ambiguity of DOA estimation by leveraging the IRS
illumination diversity across two phases. Therefore, it works well
even for rank-one AP-IRS channel scenarios.

Note that our proposed algorithm can be readily adapted to the
scenario where the BS-IRS channel has a rank greater than one. In
fact, in such a case, a single tensor alone can identify the DOA
parameter. Specifically, we can resort to (\ref{corr}) to estimate
the DOA parameter when $\gamma=0$dB. From Fig.\ref{Comp_T1}, we
see that our proposed method not only presents a significant
performance improvement over the work \cite{10138058} for
approximately rank-one scenarios (corresponding to $\gamma=5$dB
and $\gamma=13$dB), but also achieves a performance close to the
work \cite{10138058} when the AP-IRS channel consists of multiple
strong paths (corresponding to $\gamma=0$dB).

\section{Conclusion}
\label{Conclusion} In this paper, we explored an IRS-assisted NLOS
sensing system. We introduced a radar operation mode for the AP,
which senses the NLOS environment by transmitting OFDM pulses and
processing echoes relayed by the IRS. A two-phase sensing scheme
was proposed by exploiting the diversity in the illumination
pattern of the IRS across two different phases. Using this
two-phase sensing approach, we developed a CP decomposition-based
method for estimating the DOA, Doppler shifts, and time delays of
the targets. Uniqueness conditions for the proposed method are
analyzed and provided. We also conducted a CRB analysis for the
considered estimation problem. Simulation results demonstrated the
effectiveness of the proposed method in performing NLOS sensing,
even in scenarios where there was only a single dominant path
between the AP and the IRS.

\bibliography{refs}

\begin{thebibliography}{10}
\providecommand{\url}[1]{#1}
\csname url@samestyle\endcsname
\providecommand{\newblock}{\relax}
\providecommand{\bibinfo}[2]{#2}
\providecommand{\BIBentrySTDinterwordspacing}{\spaceskip=0pt\relax}
\providecommand{\BIBentryALTinterwordstretchfactor}{4}
\providecommand{\BIBentryALTinterwordspacing}{\spaceskip=\fontdimen2\font plus
\BIBentryALTinterwordstretchfactor\fontdimen3\font minus \fontdimen4\font\relax}
\providecommand{\BIBforeignlanguage}[2]{{%
\expandafter\ifx\csname l@#1\endcsname\relax
\typeout{** WARNING: IEEEtran.bst: No hyphenation pattern has been}%
\typeout{** loaded for the language `#1'. Using the pattern for}%
\typeout{** the default language instead.}%
\else
\language=\csname l@#1\endcsname
\fi
#2}}
\providecommand{\BIBdecl}{\relax}
\BIBdecl

\bibitem{8811733}
Q.~Wu and R.~Zhang, ``Intelligent reflecting surface enhanced wireless network via joint active and passive beamforming,'' \emph{IEEE Transactions on Wireless Communications}, vol.~18, no.~11, pp. 5394--5409, 2019.

\bibitem{9140329}
M.~Di~Renzo, A.~Zappone, M.~Debbah, M.-S. Alouini, C.~Yuen, J.~de~Rosny, and S.~Tretyakov, ``Smart radio environments empowered by reconfigurable intelligent surfaces: How it works, state of research, and the road ahead,'' \emph{IEEE Journal on Selected Areas in Communications}, vol.~38, no.~11, pp. 2450--2525, 2020.

\bibitem{8910627}
Q.~Wu and R.~Zhang, ``Towards smart and reconfigurable environment: Intelligent reflecting surface aided wireless network,'' \emph{IEEE Communications Magazine}, vol.~58, no.~1, pp. 106--112, 2020.

\bibitem{9326394}
Q.~Wu, S.~Zhang, B.~Zheng, C.~You, and R.~Zhang, ``Intelligent reflecting surface-aided wireless communications: A tutorial,'' \emph{IEEE Transactions on Communications}, vol.~69, no.~5, pp. 3313--3351, 2021.

\bibitem{9226616}
P.~Wang, J.~Fang, X.~Yuan, Z.~Chen, and H.~Li, ``Intelligent reflecting surface-assisted millimeter wave communications: Joint active and passive precoding design,'' \emph{IEEE Transactions on Vehicular Technology}, vol.~69, no.~12, pp. 14\,960--14\,973, 2020.

\bibitem{9234098}
P.~Wang, J.~Fang, L.~Dai, and H.~Li, ``Joint transceiver and large intelligent surface design for massive mimo mmwave systems,'' \emph{IEEE Transactions on Wireless Communications}, vol.~20, no.~2, pp. 1052--1064, 2021.

\bibitem{9927151}
B.~Ning, P.~Wang, L.~Li, Z.~Chen, and J.~Fang, ``Multi-irs-aided multi-user mimo in mmwave/thz communications: A space-orthogonal scheme,'' \emph{IEEE Transactions on Communications}, vol.~70, no.~12, pp. 8138--8152, 2022.

\bibitem{9540344}
J.~A. Zhang, F.~Liu, C.~Masouros, R.~W. Heath, Z.~Feng, L.~Zheng, and A.~Petropulu, ``An overview of signal processing techniques for joint communication and radar sensing,'' \emph{IEEE Journal of Selected Topics in Signal Processing}, vol.~15, no.~6, pp. 1295--1315, 2021.

\bibitem{9737357}
F.~Liu, Y.~Cui, C.~Masouros, J.~Xu, T.~X. Han, Y.~C. Eldar, and S.~Buzzi, ``Integrated sensing and communications: Toward dual-functional wireless networks for 6g and beyond,'' \emph{IEEE Journal on Selected Areas in Communications}, vol.~40, no.~6, pp. 1728--1767, 2022.

\bibitem{9893187}
H.~Wymeersch and G.~Seco-Granados, ``Radio localization and sensing—part i: Fundamentals,'' \emph{IEEE Communications Letters}, vol.~26, no.~12, pp. 2816--2820, 2022.

\bibitem{9893114}
------, ``Radio localization and sensing—part ii: State-of-the-art and challenges,'' \emph{IEEE Communications Letters}, vol.~26, no.~12, pp. 2821--2825, 2022.

\bibitem{9508883}
A.~Aubry, A.~De~Maio, and M.~Rosamilia, ``Reconfigurable intelligent surfaces for n-los radar surveillance,'' \emph{IEEE Transactions on Vehicular Technology}, vol.~70, no.~10, pp. 10\,735--10\,749, 2021.

\bibitem{9454375}
S.~Buzzi, E.~Grossi, M.~Lops, and L.~Venturino, ``Radar target detection aided by reconfigurable intelligent surfaces,'' \emph{IEEE Signal Processing Letters}, vol.~28, pp. 1315--1319, 2021.

\bibitem{9732186}
------, ``Foundations of mimo radar detection aided by reconfigurable intelligent surfaces,'' \emph{IEEE Transactions on Signal Processing}, vol.~70, pp. 1749--1763, 2022.

\bibitem{9361184}
W.~Lu, Q.~Lin, N.~Song, Q.~Fang, X.~Hua, and B.~Deng, ``Target detection in intelligent reflecting surface aided distributed mimo radar systems,'' \emph{IEEE Sensors Letters}, vol.~5, no.~3, pp. 1--4, 2021.

\bibitem{9647914}
F.~Wang, H.~Li, and J.~Fang, ``Joint active and passive beamforming for irs-assisted radar,'' \emph{IEEE Signal Processing Letters}, vol.~29, pp. 349--353, 2022.

\bibitem{10149462}
Z.~Esmaeilbeig, A.~Eamaz, K.~V. Mishra, and M.~Soltanalian, ``Moving target detection via multi-irs-aided ofdm radar,'' in \emph{2023 IEEE Radar Conference (RadarConf23)}, 2023, pp. 1--6.

\bibitem{9724202}
X.~Shao, C.~You, W.~Ma, X.~Chen, and R.~Zhang, ``Target sensing with intelligent reflecting surface: Architecture and performance,'' \emph{IEEE Journal on Selected Areas in Communications}, vol.~40, no.~7, pp. 2070--2084, 2022.

\bibitem{9827797}
Z.~Esmaeilbeig, K.~V. Mishra, and M.~Soltanalian, ``Irs-aided radar: Enhanced target parameter estimation via intelligent reflecting surfaces,'' in \emph{2022 IEEE 12th Sensor Array and Multichannel Signal Processing Workshop (SAM)}, 2022, pp. 286--290.

\bibitem{10138058}
X.~Song, J.~Xu, F.~Liu, T.~X. Han, and Y.~C. Eldar, ``Intelligent reflecting surface enabled sensing: Cramér-rao bound optimization,'' \emph{IEEE Transactions on Signal Processing}, vol.~71, pp. 2011--2026, 2023.

\bibitem{9937163}
Z.~Yu, X.~Hu, C.~Liu, M.~Peng, and C.~Zhong, ``Location sensing and beamforming design for irs-enabled multi-user isac systems,'' \emph{IEEE Transactions on Signal Processing}, vol.~70, pp. 5178--5193, 2022.

\bibitem{10141975}
C.~Liao, F.~Wang, and V.~K.~N. Lau, ``Optimized design for irs-assisted integrated sensing and communication systems in clutter environments,'' \emph{IEEE Transactions on Communications}, vol.~71, no.~8, pp. 4721--4734, 2023.

\bibitem{10130707}
E.~Shtaiwi, H.~Zhang, A.~Abdelhadi, A.~L. Swindlehurst, Z.~Han, and H.~V. Poor, ``Sum-rate maximization for ris-assisted integrated sensing and communication systems with manifold optimization,'' \emph{IEEE Transactions on Communications}, vol.~71, no.~8, pp. 4909--4923, 2023.

\bibitem{10422881}
X.~Shao, C.~You, and R.~Zhang, ``Intelligent reflecting surface aided wireless sensing: Applications and design issues,'' \emph{IEEE Wireless Communications}, pp. 1--7, 2024.

\bibitem{5393298}
C.~R. Berger, B.~Demissie, J.~Heckenbach, P.~Willett, and S.~Zhou, ``Signal processing for passive radar using ofdm waveforms,'' \emph{IEEE Journal of Selected Topics in Signal Processing}, vol.~4, no.~1, pp. 226--238, 2010.

\bibitem{9420261}
M.~F. Keskin, V.~Koivunen, and H.~Wymeersch, ``Limited feedforward waveform design for ofdm dual-functional radar-communications,'' \emph{IEEE Transactions on Signal Processing}, vol.~69, pp. 2955--2970, 2021.

\bibitem{harshman1972determination}
R.~A. Harshman, ``Determination and proof of minimum uniqueness conditions for parafac1,'' \emph{UCLA working papers in phonetics}, vol.~22, no. 111-117, p.~3, 1972.

\bibitem{kruskal1977three}
J.~B. Kruskal, ``Three-way arrays: rank and uniqueness of trilinear decompositions, with application to arithmetic complexity and statistics,'' \emph{Linear algebra and its applications}, vol.~18, no.~2, pp. 95--138, 1977.

\bibitem{stegeman2007kruskal}
A.~Stegeman and N.~D. Sidiropoulos, ``On kruskal’s uniqueness condition for the candecomp/parafac decomposition,'' \emph{Linear Algebra and its applications}, vol. 420, no. 2-3, pp. 540--552, 2007.

\bibitem{6573422}
M.~Sørensen and L.~De~Lathauwer, ``Blind signal separation via tensor decomposition with vandermonde factor: Canonical polyadic decomposition,'' \emph{IEEE Transactions on Signal Processing}, vol.~61, no.~22, pp. 5507--5519, 2013.

\bibitem{9049103}
Y.~Lin, S.~Jin, M.~Matthaiou, and X.~You, ``Tensor-based channel estimation for millimeter wave mimo-ofdm with dual-wideband effects,'' \emph{IEEE Transactions on Communications}, vol.~68, no.~7, pp. 4218--4232, 2020.

\bibitem{7914672}
Z.~Zhou, J.~Fang, L.~Yang, H.~Li, Z.~Chen, and R.~S. Blum, ``Low-rank tensor decomposition-aided channel estimation for millimeter wave mimo-ofdm systems,'' \emph{IEEE Journal on Selected Areas in Communications}, vol.~35, no.~7, pp. 1524--1538, 2017.

\bibitem{10274514}
P.~Wang, W.~Mei, J.~Fang, and R.~Zhang, ``Target-mounted intelligent reflecting surface for joint location and orientation estimation,'' \emph{IEEE Journal on Selected Areas in Communications}, vol.~41, no.~12, pp. 3768--3782, 2023.

\bibitem{6834753}
M.~R. Akdeniz, Y.~Liu, M.~K. Samimi, S.~Sun, S.~Rangan, T.~S. Rappaport, and E.~Erkip, ``Millimeter wave channel modeling and cellular capacity evaluation,'' \emph{IEEE Journal on Selected Areas in Communications}, vol.~32, no.~6, pp. 1164--1179, 2014.

\bibitem{8386686}
S.~Sun, T.~S. Rappaport, M.~Shafi, P.~Tang, J.~Zhang, and P.~J. Smith, ``Propagation models and performance evaluation for 5g millimeter-wave bands,'' \emph{IEEE Transactions on Vehicular Technology}, vol.~67, no.~9, pp. 8422--8439, 2018.

\bibitem{6133581}
E.~Ben-Dor, T.~S. Rappaport, Y.~Qiao, and S.~J. Lauffenburger, ``Millimeter-wave 60 ghz outdoor and vehicle aoa propagation measurements using a broadband channel sounder,'' in \emph{2011 IEEE Global Telecommunications Conference - GLOBECOM 2011}, 2011, pp. 1--6.

\bibitem{7306533}
Z.~Gao, L.~Dai, D.~Mi, Z.~Wang, M.~A. Imran, and M.~Z. Shakir, ``Mmwave massive-mimo-based wireless backhaul for the 5g ultra-dense network,'' \emph{IEEE Wireless Communications}, vol.~22, no.~5, pp. 13--21, 2015.

\bibitem{9625066}
M.~Lübke, J.~Fuchs, A.~Dubey, H.~Hamoud, F.~Dressler, R.~Weigel, and F.~Lurz, ``Validation and analysis of the propagation channel at 60 ghz for vehicular communication,'' in \emph{2021 IEEE 94th Vehicular Technology Conference (VTC2021-Fall)}, 2021, pp. 1--7.

\end{thebibliography}
\bibliographystyle{IEEEtran}

\end{document}